\documentclass[conference,compsoc]{IEEEtran}

\usepackage{tikz}
\usetikzlibrary{plotmarks}
\usepackage{amsmath,amssymb}

\usepackage[english]{babel}
\usepackage{graphicx,subfigure}
\usepackage{graphics}

\usepackage{amsmath}
\usepackage[cmintegrals]{newtxmath}
\usepackage{url}
\usepackage{listings}
\usepackage[square, numbers]{natbib}
\usepackage{xcolor}
\usepackage{xspace}
\usepackage{booktabs}
\usepackage{tablefootnote}
\usepackage{algorithm}
\usepackage{algorithmic}
\usepackage{adjustbox}
\usepackage{comment}
\usepackage{array}
\usepackage{longtable}

\expandafter\def\expandafter\UrlBreaks\expandafter{\UrlBreaks
  \do\a\do\b\do\c\do\d\do\e\do\f\do\g\do\h\do\i\do\j%
  \do\k\do\l\do\m\do\n\do\o\do\p\do\q\do\r\do\s\do\t%
  \do\u\do\v\do\w\do\x\do\y\do\z\do\A\do\B\do\C\do\D%
  \do\E\do\F\do\G\do\H\do\I\do\J\do\K\do\L\do\M\do\N%
  \do\O\do\P\do\Q\do\R\do\S\do\T\do\U\do\V\do\W\do\X%
  \do\Y\do\Z}

\makeatletter
\def\lst@makecaption{%
  \def\@captype{table}%
  \@makecaption
}
\makeatother

\newcommand{\easylist}{\texttt{EASYLIST}\xspace}
\newcommand{\ublock}{\texttt{UBLOCK}\xspace}
\newcommand{\adguard}{\texttt{ADGUARD}\xspace}
\newcommand{\ats}{\texttt{ATS}\xspace}
\newcommand\name{\textsc{SINBAD}\xspace}

\newcommand\eg{\emph{e.g.},\xspace}
\newcommand\ie{\emph{i.e.},\xspace}
\newcommand\etc{\emph{etc}.\xspace}

\providecommand{\etal}{\emph{et al.}\xspace}

\newcommand{\broken}{\texttt{Broken}\xspace}
\newcommand{\legit}{\texttt{Legitimate}\xspace}
\newcommand{\neutral}{\texttt{Neutral}\xspace}

\newcommand{\para}[1]{\smallskip \noindent \textbf{#1}}
\newcommand{\parait}[1]{\smallskip \noindent \textit{#1}}

\begin{document}
\title{\name: Saliency-informed detection of breakage caused by ad blocking}

\author{\IEEEauthorblockN{Saiid El Hajj Chehade}
\IEEEauthorblockA{EPFL}
\and
\IEEEauthorblockN{Sandra Siby}
\IEEEauthorblockA{Imperial College London}
\and
\IEEEauthorblockN{Carmela Troncoso}
\IEEEauthorblockA{EPFL}}

\maketitle

\begin{abstract}
Privacy-enhancing blocking tools based on filter-list rules tend to break legitimate functionality. Filter-list maintainers could benefit from automated breakage detection tools that allow them to proactively fix problematic rules before deploying them to millions of users.
We introduce \name, an automated breakage detector that improves the accuracy over the state of the art by 20\%, and is the first to detect dynamic breakage and breakage caused by style-oriented filter rules.
The success of \name is rooted in three innovations: (1) the use of user-reported breakage issues in forums that enable the creation of a high-quality dataset for training in which only breakage that users perceive as an issue is included; (2) the use of `web saliency' to automatically identify user-relevant regions of a website on which to prioritize automated interactions aimed at triggering breakage; and (3) the analysis of web-pages via subtrees which enables fine-grained identification of problematic filter rules. 
\end{abstract}

\section{Introduction}
\label{sec:introduction}
Privacy-enhancing blocking tools~\cite{adblockplus_web, ubo_web, ghostery_web,brave_tracking_protection, edge_tracking_protection} operate either by blocking network requests or by hiding elements rendered on a webpage.
Blocking tools primarily rely on filter lists (\eg EasyList~\cite{easylist} or EasyPrivacy~\cite{easyprivacy}) that are manually curated by a small community of maintainers.
These filter lists contain rules that describe which resources should be blocked or hidden during a web-page load. 

While applying these rules, blocking tools can cause legitimate parts of a webpage to stop functioning, a phenomenon commonly known as \textit{breakage}. 
Breakage causes negative user experience, affects adoption of blocking tools~\cite{mathur2018characterizing, nisenoff2023defining}, and prevents the tools' developers from adopting aggressive blocking policies, reducing the protection these tools could provide~\cite{sahib2022bringing}.

Filter-list maintainers 
typically fix breakage upon reports from users, in a slow and burdensome process. 
This process could be automatized to detect when changes in filter lists' rules cause breakage and revert those changes before those updates impact millions of users.
However, automatic detection of breakage is very challenging for two reasons. 
First, it is hard to automatically trigger breakage induced by user interactions, \eg a video does that does not play; and second, breakage detection has a subjective component: a blocked video ad and a blocked legitimate video exhibit technically the same page behavior, but cause a different user experience:. 
Thus, it is difficult to collect breakage samples that can be used to train breakage detectors. Moreover, breakage is often hard to reproduce due to legitimate changes on the website, 3rd-party APIs, and URL paths.

In this paper, we introduce a pipeline that enables training of machine-learning-based breakage detectors.
We use this pipeline to build \name, a breakage detector that uses web `saliency' -- which is a proxy for the importance of elements within a webpage -- to prioritize interactions.
To account for subjectivity, we extract breakage instances from ad-blocking forums where users report breakage issues; and to avoid including breakage caused by other causes than filter-list rules, we only include in the training issues where we have evidence that the issue generated a filter-list fix. 

Our contributions are as follows:

\begin{itemize}
    \item We build a high-quality dataset for breakage detection from user-reported breakage issues on forums. We find that breakage reports take from days to weeks to be resolved, highlighting the importance of automated breakage detection tools that allow maintainers to be proactive. We also find that dynamic breakage corresponds to 25\% of breakage issues and that CSS-hiding filter-list rules are a $\approx 53$ \% of rules causing breakage on average, neither of which are covered by state-of-the-art detectors~\cite{smith2022blocked}.
    \item We show that it is possible to automatically identify important, `salient', regions on a webpage, and prioritizing automated interactions in these regions enables the discovery of user-relevant breakage with much less effort than random interactions.
    \item We propose \name, a saliency-informed breakage detection system that identifies breakage with 20\% better accuracy than the state of the art. \name correctly classifies breakage that previous approaches miss by design, such as dynamic breakage after user interactions, and content breakage stemming from CSS-based filter rules. \name identifies broken regions of a page instead of classifying entire pages, enabling fine-grained fixing of blocking rules.
\end{itemize}

\section{Background \& Related Work}
\label{sec:background}

\label{subsec:adblockers}

Many ad and tracking services (\ats) blocking tools  rely on manually-curated filter lists (\eg ~\cite{easylist,easyprivacy,disconnect_tracking_list}) to block \ats, \eg AdBlock Plus~\cite{adblockplus_web}, uBlock Origin~\cite{ubo_web}, Ghostery~\cite{ghostery_web}, AdGuard~\cite{AdGuardAdBlockerBrowser}, extensions or in-browser protections in browsers such as Firefox~\cite{firefox_cookie_policy}, Edge~\cite{edge_tracking_protection}, and Brave~\cite{brave_tracking_protection}. 
Maintainers of ad-blocking tools typically rely on manual (often, visual) verification of a small subset of websites to determine whether blocking causes loss of legitimate functionality on websites (\textit{breakage}) that could lead to negative user experience. 
To address the scalability and robustness issues of manual curation ~\cite{iqbal17antiABIMC, snyder20whofilters,wang16webranz, alrizah19errorsfilterlists,le2022cvinspector}, the ad-blocking research community has proposed machine-learning approaches to automate the detection of \ats~\cite{iqbal20adgraph, iqbal21fingerprinting, iqbal22khaleesi, siby22webgraph, yang2022wtagraph, munir23cookiegraph}. 

Neither manual nor automated \ats-detection approaches include extensive checks to proactively identify rules that cause breakage. Maintainers address breakage issues in a reactive manner when users of ad-blocking tools report breakage~\cite{adblockplus_forum, ubo_issues, adguard_issues}. Maintainers replicate the breakage issue via manual checks, find the filter-list rule(s) that caused the issue, and update the(se) rule(s).
 
Our work complements existing \ats detection tools by enabling maintainers to determine, \textit{proactively} and \textit{at scale}, whether new or fixed filter-list rules cause breakage. 

\para{User studies on breakage.} 
Mathur \etal~\cite{mathur2018characterizing} investigate user attitudes towards online tracking and the measures users took to protect themselves.
Their study found that breakage was relatively uncommon.
Nisenoff \etal~\cite{nisenoff2023defining} find a higher prevalence of breakage and propose a taxonomy of user experiences of breakage by analyzing public user reviews and issue reports of popular blocking tools. 
In our work, we use breakage issue reports as a source of training data, considering only reports for which there is evidence that breakage occurred due to a filter-list rule.

\para{Heuristics-based breakage detection.} 
Previous works quantify breakage via various heuristics either in place of, or in addition to, manual checks.
Krishnamurthy \etal~\cite{krishnamurthy2007measuring} use the number of visual elements on a page as a metric to calculate page quality;
Yu \etal~\cite{yu2016tracking} measure how often users reloaded a page; 
Jueckstock \etal~\cite{jueckstock2022measuring} quantify the similarity in edge sets between graph representations of a page with and without a policy applied to it;
Fouquet \etal~\cite{fouquet2023breaking} use heuristics based on analyzing documentation and common knowledge of practices in the field;
Le \etal~\cite{le2022autofr} detect visual breakage by comparing the number of visible non-ad images and text before and after applying an intervention; 
Amjad \etal~\cite{amjad2023blocking} use the differences in the number of functional HTTP requests and HTML tags with functional src attributes;
Castell-Uroz \etal~\cite{castell2023astrack} measure visual differences of website screenshots with and without blocking, and manually verify the results.

All the above approaches cover only \textit{static} breakage that does not require user interaction to trigger it. 
Yet, from 25 up to 44\% of user-reported examples of breakage are dynamic (see Section~\ref{sec:dataset}) and cannot be detected by these approaches.
In addition, metrics based on counting the number of visual elements, requests, or tags are influenced by webpages' natural dynamism -- pages can fetch different numbers and types of resources, leading to variations in these counts -- which leads to errors in detection.
We show that \name outperforms all these approaches in Section~\ref{subsec:compare-methods}. 

\para{Machine-learning-based breakage detection.}
To the best of our knowledge, there exists only one machine-learning-based breakage detection approach. Smith \etal~\cite{smith2022blocked} build a classifier trained on graph representations of page-load events and use the EasyList commit history to label data. 
We show that using the commit history can result in incorrect ground-truth labeling; and that \name outperforms Smith \etal's approach by 20\%. Furthermore, \name correctly classifies dynamic breakage that is missed by design in\cite{smith2022blocked}.

\para{Other approaches to avoid breakage} Existing alternatives to avoid breakage, such as automatically replacing tracking JavaScript code with alternatives that preserve functionality~\cite{smith2021sugarcoat}, are hard to deploy due to scalability issues.

\section{Obtaining Breakage Examples}
\label{sec:dataset}

In this section, we describe the methodology to collect  user-reported breakage issues from public ad-blocker forums and determine whether they are suitable as training data for \name.
The main characteristics we consider are (1) \textit{validity}, \ie a post represents a breakage that is caused by a filter-list rule, (2) \textit{automatability}, \ie the post can be easily parsed to extract breakage details to enable replication, and (3) \textit{reproducibility} \ie we can recreate the reported breakage issue. 
We also discuss the limitations of existing breakage-detection systems given the classes of breakage users report.

\subsection{Data sources}

We investigate three public data sources. We select these data sources because, in all of them, maintainers link to a GitHub commit when resolving issues, giving a strong indication that breakage was caused by a filter-list rule, as opposed to errors due to other factors (\eg programming errors or slow load times).
The sources are:

\parait{EasyList.} The EasyList ``Report incorrectly removed content'' public forum is where users report issues caused by the EasyList~\cite{easylist} filter lists~\cite{easylist_forum}.
Users' posts typically include the URL of the broken webpage and a description of the experienced breakage, and may contain images illustrating this breakage. 
Users may also point to the filter-list rule that they think caused the breakage.
When breakage is due to filter-list rules, the filter-list maintainers update the filter lists and post a link to the EasyList GitHub commit of the fix.

\parait{uBlock.} The uBlock Origin's uAssets GitHub issue tracker is where users of the tool~\cite{ubo_web} report instances of breakage~\cite{ubo_issues}. 
Posts in the issue tracker have four subsections: URL, category, description, and screenshots; although users can deviate from this format.
The maintainers of the tool reply to breakage reports with a link to the uBlock GitHub commit of the fix and close the issue.

\parait{AdGuard.} The AdguardFilters GitHub issue tracker~\cite{adguard_issues} for the AdGuard ad-blocker~\cite{AdGuardAdBlockerBrowser}. 
The structure of the posts is very similar to the uBlock repository, but issues are submitted through the ad-blocker interface.
All posts have the same structure and feature the test URL first, followed by screenshots, and system configuration (which filter lists the user had installed).
The maintainers of this forum add labels to the issues, clearly identifying breakage, in addition to other relevant information for our study such as ``could not reproduce''. 
The issues on AdGuard's issue tracker are more recent than the other two sources, indicating that they are more likely to be reproducible.

\para{Sources we do not include.} 
We do not consider two data sources that have been used in previous work.
We do not use marketplace reviews of ad-blocking extensions where users might report breakage~\cite{nisenoff2023defining}, as we cannot know with certainty whether user-reported breakage was the result of a filter-list rule. 
We also do not include GitHub issues from Privacy Badger~\cite{nisenoff2023defining}, because Privacy Badger relied on heuristics instead of filter lists till October 2020. 

We also analyzed other sources in the ad-blocking space, and discarded them for various reasons: 
lack of replies that enable us to identify filter-list-related breakage~~\cite{adblockplus_forum,brave_community}; limited number of breakage examples~\cite{sitepoint}; issues being unrelated to block-related breakage (\eg tailored to compatibility across browsers~\cite{webcompat}).

\para{Ethical considerations.} All the posts and issues we analyze are from publicly-available data sources. 
We do not collect or process any identifiable information such as usernames.  
Our data collection and analysis procedure was approved by our institutional ethics board.
Prior to data collection, we informed the maintainers of the forums of our practices.

\subsection{Dataset Collection and Processing}
\label{subsec:dataset-scraping}

Our breakage issues' collection process works as follows. 
For each issue, we collect the post title, the creation timestamp, the post URL, and the (cleaned) post content. We also collect the filter lists that are most likely to have created the breakage and the filter lists that fix this breakage (we also collect the maintainer's commits to the filter-list repositories where they established the `breaking' and `fixing' filter lists).
We summarize the collected data in Table~\ref{table:datasets}.

\para{\easylist.}
We scrape the posts in the EasyList forum using Beautiful Soup~\cite{beautiful_soup}. 
We first crawl the main forum page to obtain a list of links to the issues' posts. 
From this list, we filter issues that have the \texttt{Locked} tag, indicating maintainers have addressed them. 
For each of these issues, we store the \textit{title}, the \textit{creation date}, and the \textit{post URL}.
We disregard issues that have no \textit{commit URLs} by the maintainer in the post replies.
Removing all the non-fixed issues, we obtain 7,900 breakage examples since 2006.

\para{\ublock.}
We use the GitHub API to scrape the issues in uBlock's repository (\texttt{uBlockOrigin/uAssets/issues}), filtering posts that contain the keywords ``breakage'', ``Breakage'', or ``[Breakage]'' in the title. 
We keep only closed issues (\textit{state=closed}) if they have a commit to a fix. 
We extract the \textit{title}, \textit{timestamp}, \textit{post URL}, the filter lists used by the user during the breakage (as declared in the post), and the moderators' commits to uBlock GitHub in the responses. 
We obtain 638 resolved posts over a period of 5 years.

\para{\adguard.}
We use the GitHub API to scrape AdGuard resolved breakage issues with the \texttt{T: Incorrect Blocking} and \texttt{A: Resolved} labels. 
We extract the same information as for \ublock. 
We use keyword-based heuristics (details in the appendix) to extract the URL of the broken page provided by the user and manually find the URL for issues where our heuristics fail.
We obtain 8,992 scraped fixed issues published between 2015 and 2023. 

\para{Differences with respect to Smith \etal~\cite{smith2022blocked}.} 
Smith \etal assume that all commits that update the filter list are a fix to breakage happening on the URL appearing in the message of the commit. By studying the issues in the forums, we observed that, in many cases, resolving a breakage issue often requires multiple rounds of iterative fixes based on user feedback, which is reflected in multiple commits by the maintainers (18\% of our EasyList issues). This leads to their dataset containing samples where breakage is not fully fixed, introducing noise in the classifier training.

To make sure we collect only filter lists that result in breakage and that fix it, we use the times of the posts of maintainers in the forum. We use the first response of the maintainer that has an update commit to identify the `breaking' list, assuming that the list that best approximates the one causing the breakage in the issue is the one prior to that update. To identify the `fixing' list, we use the last commit associated with the issue, assuming that when the interaction between users and maintainers stops it is because the issue is fixed. This greatly reduces the number of examples we can obtain compared to Smith \etal, but provides better guarantees that samples represent true breakage and fixing. We manually test 10\% of our samples to validate that this assumption is correct.

A second difference with respect to the approach of Smith \etal~\cite{smith2022blocked} is that in their experiments, they only use the filter-list rules altered in the commit, rather than the full lists.
Our manual checks reveal that such an approach results in many instances of breakage not being triggered, also resulting in noisy training data.
The reason is that in some cases breakage is caused by interdependencies between rules, instead of a particular rule. 
For example, an altered filtered rule that is an exception to a non-altered blocking rule. If used on its own, it has no effect. We illustrate this problem in Appendix~\ref{sec:interdependency}.
In our experiments, we always use the complete filter lists at the time of breakage and fixing.

\begin{table}
	\caption{Overview of breakage datasets.}
	\centering
	\resizebox*{\columnwidth}{!}
 {
	\begin{tabular}{llccc}
		\toprule
		\textbf{Dataset} & \textbf{Identifier} & \textbf{\# Scraped} & \textbf{\# Included} & \textbf{Usable Timespan} \\
		\midrule
		EasyList forum & \easylist & 7900 & 1344 & 2006-2022 \\
		uBlock Origin issues & \ublock & 638 & 543 & 2017-2022\\
            AdGuard Filters issues & \adguard & 8992 & 2605 & 2015-2023\\
		\bottomrule
	\end{tabular}
	}
	\label{table:datasets}
\end{table}

\subsection{Dataset Analysis}

\subsubsection{Need for automated breakage detection}
We first validate that fixing issues is a time-consuming process by computing the average \textit{Fix time} of the issues: the time difference between the time a user published a post reporting an issue and the time of resolution (marking the issue as fixed or closed). We identify a fixed issue as any closed issue having at least one commit from the maintainer. 
We find that the fix time varies significantly across issues and across the two datasets.
Users needs to wait, on average, 14 (Std: 194) days on average before the issue is fixed on \easylist, 36 (Std: 136) days on \ublock, and 5 (Std: 20) days on \adguard. 

\para{Takeaways.} Since the wait time for resolving breakage issues can be fairly large (in the order of months for \easylist and \ublock), it would be very beneficial to have an automated tool such as \name to test for breakage before updating filter lists publicly.

\begin{figure}[!t]
	\centering
	\includegraphics[width=\linewidth]{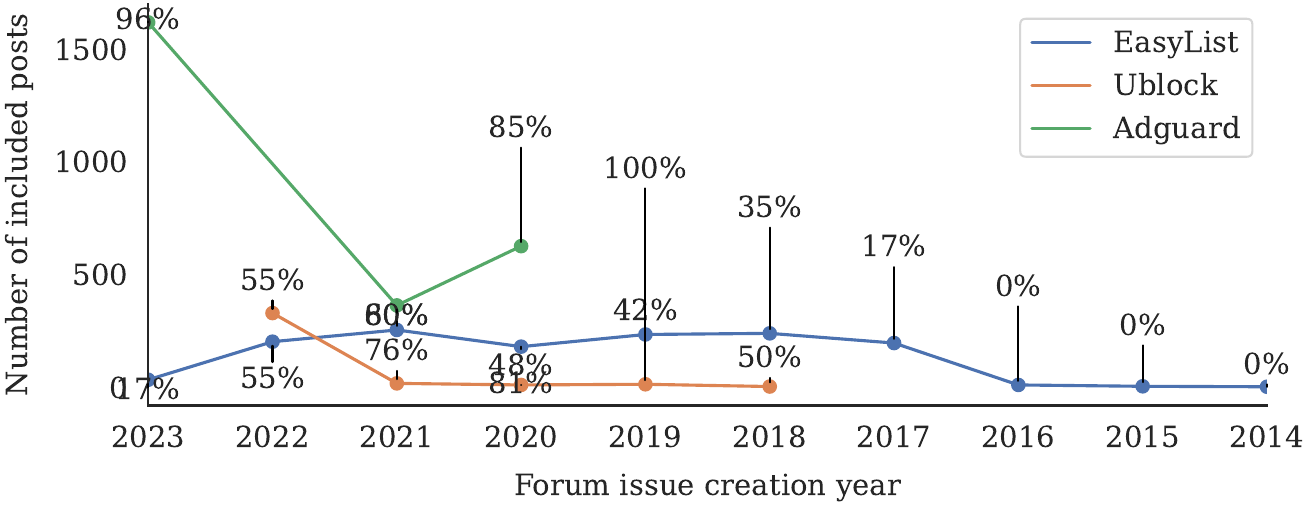}
	
 \caption[]{Number of issues that we can process and have alive test URLs, annotated with their proportion from all issues created in that year.}
    \label{fig:issue-included-time}
\end{figure}

\subsubsection{Automatability and Reproducibility}
\label{subsubsec:reproduce-analysis}
We analyze our data sources to ensure they are suitable as training data.

\para{Post structure.} 
The structure and format of posts in the forums evolved over time due to changes in forum-posting guidelines and users changing their breakage description patterns. 
We use a heuristic inspired by the most recent and frequent posting patterns to extract the information of interest described in the previous section. We also test whether the mentioned URL is still alive or not. If any of these two operations fail, we discard the post. 

We show the percentage of issues that we can properly parse per year in Figure~\ref{fig:issue-included-time}. 
For \easylist, we only consider the 1,344 issues after 2016, as we cannot parse any issue before that year. 
For \ublock, we consider all 543 issues that we can parse. 
Finally, in \adguard, we find few parsing problems, and we stop the scraping in 2020 for storage and time constraints, obtaining 2,605 posts.

\para{Reproducibility.} 
Prior work~\cite{smith2022blocked} relied on heuristics based on network activity to determine reproducibility, which we find to be inaccurate (Section~\ref{sec:evaluation}). 

We conduct a manual evaluation of breakage reproducibility to characterize the quality of our data sources. 
We follow three criteria. 
If the user specifies geographic limitations, login requirements, or other unavoidable challenges (unclear description, complex interactions required, \etc), we mark the issue as not reproducible. 
Next, we check if the site is still live and, when present, if the site matches the screenshots in the forum. If any of these fail, we mark the issue as not reproducible. 
Finally, we load the breaking filter list scraped for this issue and follow the instructions given by the user to compare our experience with their complaints and screenshots, \eg checking for missing images, unclickable buttons, scrolling impossible \etc If the results do not match the post content, we mark it as not reproducible. If the issue passes all checks, we consider it reproducible.

\begin{figure}[!t]
	\centering
	\includegraphics[width=\linewidth]{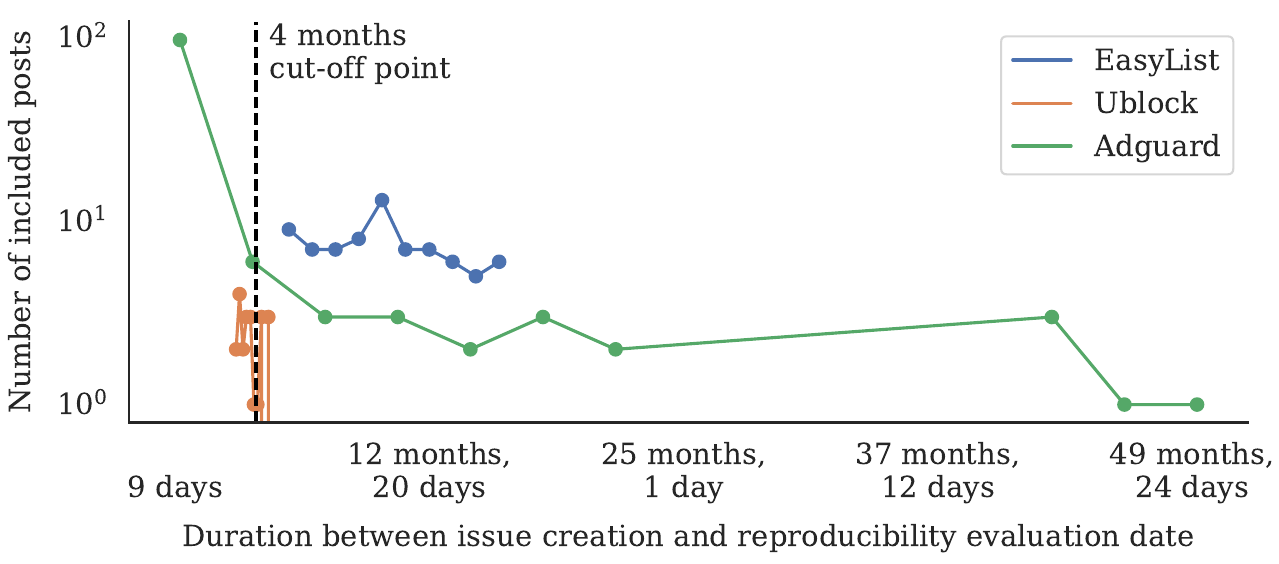}
	\caption[]{Plot showing the number of reproducible issues according to the duration between the time when the issue was created and when we evaluated whether it can be reproduced.}
    \label{fig:repr-over-time}
\end{figure}

We analyze 170 posts in \easylist (13\% of total posts), 57 posts in \ublock (10\% of total posts), and 209 posts in \adguard (8\% of total posts). 
We find that around 41\% of \easylist's issues, 38\% of \ublock's issues, and 66\% of \adguard's issues are reproducible.
This proportion decreases rapidly as the issues become older (see Figure \ref{fig:repr-over-time}). 
For \easylist, fewer than 50\% of the issues are reproducible for issues created one year before our analysis. 
For \ublock, we can only reproduce 16\% of the issues older than 5 months and only 50\% of those that were reported in the 4 months before the analysis. 
Thus, we restrict our manual analysis to 17 months for \easylist and 5 months for \ublock. In \adguard, we can reproduce more than 76\% of issues in the last 4 months -- before it drops sharply.

The main causes of non-reproducibility for the datasets are: the page being outdated/changed (25\% in \easylist and 36\% in \ublock), the domain no longer being active (7\% in \easylist and 12\% in \ublock), the page not being accessible \eg URL to a deleted blog post (4\% in \easylist and 40\% in \ublock), and the page being behind a login wall (4\% in \easylist). Other note-worthy reasons are different geographic zones, complex interaction sequences, and browser-specific issues. 

\para{Take aways.}
Our analysis shows that the utility of forums as sources for breakage research is limited by posts structure inconsistencies and the lack of reproducibility. 
For the latter, the main reason is that webpages change over time (more than 25\% of the unreproducible issues), meaning that datasets expire over time.
As a result of these problems, after removing non-parseable and non-reproducible forum issues, the size of both \easylist and \ublock datasets shrink by 98\% and 70\% respectively, from 7,900 in \easylist to just 170 and from 638 in \ublock to just 203.
Thus, in our experiments we mainly use the 512 \adguard issues from the last 4 months that we can easily parse and reproduce; and only use \easylist and \ublock for validation.

\subsubsection{Breakage characterization}

We study the the kind of reported breakage to understand the extent to which previous work can address user experiences.

\para{Filter-list rule type.}
In general, filter-list rules are categorized into two groups: \textit{blocking} rules and \textit{content} rules.

Blocking rules are applied at the network level to determine whether to block a particular network request.
Breakage caused by blocking rules can be replicated offline -- for example, we can log all the requests that occur during a page visit without an ad-blocker and then perform rule-matching to simulate the decisions of an ad-blocker on the requests.

Content rules are used to hide particular elements on a page or to insert snippets that fight complex advertising strategies on the webpage. 
To hide elements, ad blockers inject ``styling'' attributes to change how the browser renders an element (\eg changing the height to zero). 
Snippets differ among ad blockers. They might implement unique scripts to be embedded in a page or more complex blocking strategies. 
Content rules cannot be analyzed offline because they cause DOM-specific behavior and run JavaScript, which requires a running browser. 

Analyzing our datasets, we find that \easylist issues are caused by 62\% blocking, 32\% content, and 6\% mixed rules; \ublock issues are caused by 36\% blocking, 60\% content rules and 4\% mixed; and \adguard issues are caused by 26\% blocking, 58\% content, and 16\% mixed. 
The larger presence of content rules in \ublock and \adguard might be attributed to the fact that those filter lists are used in ad blockers, which support a wide variety of snippets and complex hiding techniques. 
\easylist, on the contrary, is designed to be compatible with most ad blockers, and hence uses fewer complex hiding techniques. 

\para{Prevalence of dynamic breakage.}
We also study whether breakage is  \textit{static} -- it does not require user interaction (\eg a missing video), or \textit{dynamic} -- it requires at least one user interaction such as clicking, scrolling, or typing to determine that there is a problem with the page (\eg the video section might load correctly, but pressing play will result in no outcome).
From our manual checks, we find that dynamic breakage accounts for 25\% in \easylist, 44\% in \ublock, and 26\% in \adguard.

\para{Takeaways.} Given that breakage caused by content rules accounts for a considerable portion of breakage, offline approaches that log network activity and perform post-processing to trigger breakage~\cite{smith2022blocked} are not sufficient.
To observe content-rule breakage, in contrast with previous work, \name must use an online approach in which it fetches webpages.
We also observe that a large portion of issues are due to dynamic breakage. Yet no previous work can address them.
Our training-samples collection ensures that \name accounts for dynamic breakage (Section~\ref{subsubsec:interactions}).

\section{\name: Detecting website breakage}
\label{sec:detection}
\begin{figure*}[!t]
	\centering
	\includegraphics[width=\linewidth]{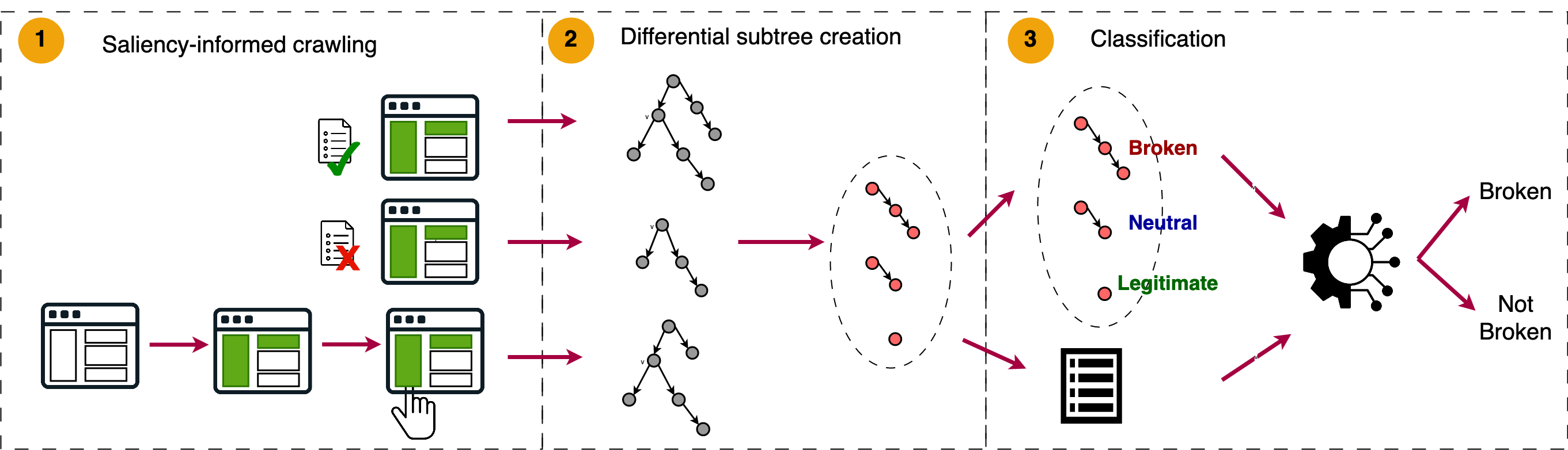}
    \label{fig:overview}
	\caption[]{Overview of \name. The pipeline consists of three phases: (1) Saliency-informed crawling: \name detects salient elements on the page, and runs three crawls -- with no filter lists, broken filter lists, and fixed filter lists. Crawls execute interactions with salient elements to trigger dynamic breakage. (2) Differential subtree creation: \name uses the changes in the page's DOM tree between pairs of crawls to create differential subtrees. (3) Subtree classification: \name extracts features and labels from the subtrees to train a classifier that can classify subtrees as broken or not.}
\end{figure*}

Figure~\ref{fig:overview} shows an overview of our automated breakage detection approach, \name. 
The \name pipeline consists of three steps.

\noindent1. \textit{Saliency-informed crawling.} \name receives a dataset of webpage URLs as input. 
It visits each pages three times -- without a filter list, with the breaking version of the list, and with the fixed version of the list. In each visit, it executes interactions with elements that are core to the user experience identified by a \emph{saliency detector}. Interactions enable \name to trigger dynamic breakage. 

\noindent2. \textit{Differential subtree creation.} For each page, \name uses the visit data to create an annotated version of the DOM tree. 
It updates the nodes in the DOM tree with information such as associated network requests or crawler interactions.
For each pair of visits to a page, \name extracts differential subtrees: the sections that changed between the DOM trees from the two visits.
These subtrees represent the modifications caused by the filter-list change to the DOM, requests, and interactions.

\noindent3. \textit{Subtree classification.} \name extracts content, structural, visual, and functional features from each subtree. It labels each subtree as broken or not, depending on changes undergone by the subtree and which pair of filter lists were used in the visits. \name uses the labeled subtrees to train a classifier that predicts whether a subtree reflects breakage. 

\subsection{Saliency-informed Crawling}

Dynamic breakage, which has not been considered in prior work, constitutes a large portion of the breakage that users experience (see Section~\ref{sec:dataset}).
Reproducing such breakage is challenging, greatly hindering the collection of data points to train a classifier.
To address this issue, \name performs \textit{saliency-based interactions}, interactions focused on webpage elements particularly relevant to users.

\subsubsection{Identifying Web-salient Areas}
\label{subsubsec:saliency}
Breakage is, by definition, an interruption in the expected user experience.
Users are more likely to interact with, and complain about, elements of interest on a page rather than peripheral elements. 
Collecting valuable breakage samples, thus, is intimately related to being able to predict users' areas of interest within a webpage and their interactions with them so as to trigger potential dynamic breakage of relevance for the user.
Borrowing the term from computer vision~\cite{ullah2020BriefSurveyVisual}, we refer to elements of interest for the user as ``salient'' elements.
In computer vision, salient regions refer to sections of an image or a video that catch the eye of an observer. 
In a web context, sections that catch the eye are not necessarily of interest to the user~\cite{shen2014WebpageSaliency}. 
Elements such as ads, banners, or call-to-action buttons would be labeled as \textit{salient} from a pure computer vision perspective, but they are of no interest to the user, and unlikely to be considered as breakage if they would not be rendered. Because of this, traditional saliency detection techniques cannot be directly applied to the web context~\cite{chakraborty2022PredictingVisualAttention, shen2014WebpageSaliency, grier2007HowUsersView}. 

In this paper, we define \textit{web-salient} areas of a webpage as \textit{groups of DOM nodes that are an essential component to fulfill the purpose of a user's visit to the web page}.
Existing approaches to webpage-saliency detection fall into three categories. 
First, those that use DOM-structure features, \eg number of children of a node, tree depth, number of \texttt{<img>} tags, \etc \cite{vidyapu2020InvestigatingModelingWeb, utiu2018LearningWebContent, wang2023SCIEntSemanticFeatureBasedFramework}. 
Second, those that use features obtained from webpage screenshots and image data and classify on a pixel level~\cite{pang2016DirectingUserAttention, zheng2018TaskdrivenWebpageSaliency, shan2017WebpageImageSaliency}.
Third, hybrid approaches~\cite{burget2009WebPageElement} that attempt to remediate the fact that the DOM alone may fail to capture the presentation of a webpage to the user, \eg in pages where styling rules denote positions of specific nodes on the screen.
Besides using screenshots as visual features, some methods also use CSS styling like background color, font size, \etc \cite{li2015ExtractingNewsInformation}.

\para{\name's hybrid salient-areas classifier.}
In \name, we implement a hybrid approach based on structural features from the DOM and visual features from CSS styling. 

We opt to not take a vision-oriented approach based on screenshots for two reasons.
First, in \name, saliency detection is part of a crawl.
As we aim to visit as many pages as possible, we need to minimize each page's processing time and storage.
Deep learning vision models typically used for saliency map prediction, like CNNs (Convolutional Neural Networks), FCNNS (Full CNNs), and RCNNS (Residual CNNs), have high computation and storage overheads~\cite{ullah2020BriefSurveyVisual}. 
In addition, vision-oriented models output smooth saliency maps, which we would need to map to DOM elements, increasing the processing time.
In contrast, approaches based on DOM and CSS only need to process and store a small number of expert-defined features and use simpler, interpretable models that require little computation, \eg random forests or XGBoost.
They also allow us to label the group of DOM nodes directly without transformations.

To train our saliency classifier, we need a labeled dataset of salient and non-salient parts of a webpage. 
To this end, we need to first, \textit{segment webpages} into blocks; second, \textit{label those segments}; and third, \textit{extract features} for training.

\parait{Segmentation.}
Prior to detecting salient areas, we need to segment the webpage into semantic blocks.
Semantic segmentation is the process of grouping HTML nodes to form a semantic block with a \textit{meaning} to the user, \eg grouping the text fields and the button nodes of a login form. 

Web segmentation is a longstanding active research field~\cite{kiesel2021EmpiricalComparisonWeb}.%
Despite the development of many approaches, including those based on deep-learning~\cite{chen2019MMDetectionOpenMMLab, meier2017FullyConvolutionalNeural}, Kiesel \etal showed that VIPS~\cite{cai2003VIPSVisionbasedPage}, a simple rule-based method, is equal to or better than recent approaches in terms of segmentation granularity and efficiency ~\cite{kiesel2021EmpiricalComparisonWeb}. 
VIPS is a top-down heuristic algorithm that iteratively divides a webpage into a hierarchy of blocks~\cite{cai2003VIPSVisionbasedPage}. 
Blocks are divided based on DOM features and visual cues (CSS attributes, position features, fonts, \etc).  
In every iteration, VIPS subdivides blocks further. 
The number of allowed iterations is a hyperparameter that controls the granularity of the block hierarchy.

To account for the shifts in web design since VIPS's inception, we fine-tune a VIPS Python implementation~\cite{wushuartgaro2023VipsPython} to include features introduced in HTML5, \eg iframes and media-oriented features. More details about our changes can be found in Appendix~\ref{subsubsec:vips-implementation}.
Figure~\ref{fig:vips-seg}, left, shows an example of VIPS segmentation of a webpage into semantic groups (red rectangles).

\parait{Labeling.}
The datasets used in previous work are not suitable for our purpose. They either used features inaccessible to us (\eg eye-tracking data), output formats that are different from our modeling task (\eg drawing a pixel-by-pixel saliency heatmap)~\cite{shen2014WebpageSaliency, pang2016DirectingUserAttention}, or were kept private \cite{kiesel2021EmpiricalComparisonWeb, vidyapu2020InvestigatingModelingWeb}. 
Thus, we curate our own dataset. 
Following Kiesel \etal~\cite{kiesel2021EmpiricalComparisonWeb}, we select 1K websites from Alexa's top 1M sites, consisting of the top 100 sites and 900 randomly sampled sites. 
We remove web pages that have fewer than 64 elements~\cite{kiesel2020WebPageSegmentationa}, which leaves us with 543 sites.
We split the sites into five batches, and have two volunteer annotators per batch.
The annotators returned labels on 74\% of the websites. 
The rest were either reported by annotators as unusable or skipped by mistake. 
We obtain a maximum 65\% and mean 55\% Krippendorff alpha agreement measure~\cite{castro-2017-fast-krippendorff} between pairs of annotators. To address the weak inter-annotator agreement, we only consider groups considered salient by both annotators as salient. 
At the end of the process, we obtain 329 salient blocks, agreed on by both annotators, and 3,268 negative blocks. 
This imbalance is expected, as salient elements are, by definition, a small percentage of regions on a webpage.

\parait{Feature extraction.}
We compute features in four categories: \textit{structural}, \eg number of nodes in a semantic group; \textit{content} \eg the number of \texttt{<img>} tags; \textit{positional}, \eg (x, y) coordinates; and \textit{visual}, \eg color contrast. As Smith \etal~\cite{smith2022blocked}, we rank the features according to \textit{Leave-One-Covariate-Out}~\cite{lei2018DistributionFreePredictiveInferencea} referred to as AUC Loss. We compare the AUC of a modified dataset, leaving one feature out, with the AUC of the original dataset.
Higher AUC loss implies greater feature importance. 
Although this metric does not capture the relationships between features, it provides a basic understanding of the feature's impact on the model. We find a balance between presentation-specific features (\textit{positional} and \textit{visual}) and DOM-specific features (mainly \textit{content} features) between the top features, suggesting that a hybrid representation is important to predict saliency. We report the complete list of features and their predictive power in Table~\ref{tab:saliency_feature_importance}.

\parait{Model architecture and results.}
We use a random forest classifier with 100 estimators. 
To account for the imbalance in our dataset, we use SMOTE~\cite{chawla2002SMOTESyntheticMinority} to over-sample the salient minority class. 
Our classifier achieves a mean $83\pm0.05$\% AUC, and $62\pm0.09$\% F1 score for the salient class over a 5-fold cross-validation.
We verify these results by visually inspecting 10 pages.
Due to the output differences between our approach and other works~\cite{vidyapu2020InvestigatingModelingWeb,zheng2018TaskdrivenWebpageSaliency,shan2017WebpageImageSaliency}, direct comparison is not possible, but our F1 and AUC are consistent with prior works' reported performance.

\subsubsection{Interacting with salient elements}
\label{subsubsec:interactions}

Once we identify salient blocks, we need to determine which type of interactions should we perform on the elements in those blocks to trigger breakage.
Additionally, we need to develop a method to collect relevant information resulting from the interaction.

We call interaction an action sequence performed on a target element. 
We pre-define a set of actions and potential target element types. Then, during crawls, we search among the salient elements for an appropriate target. 
We consider two interactions: \textit{Typing} and \textit{Click}.
For a Typing interaction, the action sequence consists of clicking on the target, typing a random sequence of characters, and hitting enter.
Viable target candidates for typing interactions are\texttt{<textarea/>} or a text input field.
In our proof-of-concept, we use only these two interaction types, but \name's extensible design allows for new and more complex interactions to be easily integrated based on the maintainers' needs.

In addition to the network requests and page-content changes that we collect during the visit, we capture JavaScript runtime errors thrown after an interaction.
This allows us to capture dynamic breakage triggered when one or more of these scripts are blocked.
For example, dependent scripts, that rely on variables defined in blocked scripts, would raise a \texttt{ReferenceError}, and capturing these errors may help to detect this breakage.

\subsubsection{Crawl implementation}
\label{subsubsec:crawl-implementation}
We use OpenWPM v0.20.0~\cite{englehardt2016census} to automatically crawl websites with Firefox 100.0.
We augment OpenWPM with the commands to: install the ad-blocker, load the filter lists from files, dump the DOM data and salient nodes, and perform interactions on salient blocks. 

Prior work~\cite{smith2022blocked} ran two visits per website -- with and without the filter-list rules that contribute to breakage.
They then created a graph that captures the changes between the graphs obtained in the visits. 
These differences may include, on top of the broken change in the page, ads that are legitimately blocked by the broken filter list.

In our work, we run three visits per website: a visit using the filter lists resulting in breakage, ($C_{B}$), a visit using the fixed versions of filter lists ($C_{F}$), and a visit without any filter lists ($C_{N}$). Before each visit, we reset the browser to keep the visits independent.
Running three visits enables us to reduce the number of false positives with respect to previous work.
We perform $C_{F}$ first, assuming that visiting the webpage with a fixed list will contain all relevant functionality for a user and have the least number of ads.
During $C_{F}$, we identify salient regions and execute interactions on these regions.
We repeat the interactions during $C_{B}$ and $C_{N}$.
By the end of the visit, in addition to network requests and JavaScript calls, we log data previously unaccounted for in OpenWPM. This includes the DOM tree representation of the page. We also store the HTML \textit{attributes} and \textit{visual cues} of nodes in this tree. \textit{Visual cues} include data like the position on the screen, the dimensions, the text content of an element, the background color, and font size. We also store interaction timestamps, their targets, and JavaScript errors.

After running the \name pipeline on the \adguard dataset, we find that saliency reduced the interaction candidate search space from 40 elements on average per website to 2 to 3 elements per website (less than 6\%). We also argue, in Section \ref{sec:evaluation}, that saliency and interactions overall provide a significant predictive contribution.

\para{Ethical Considerations.}
We designed our crawling process to minimize the likelihood of harming the websites we visit by overloading the resources or sending data that impacts the site's services during the three main stages where the crawler operates on the web page -- accepting cookie banners, collecting static data, and interacting with the page. 

To not abuse server availability, we leave at least 50 seconds between visits to the same webpage.

To avoid hamful interactions we do the following. First, to accept cookies as a real user, the crawler uses a keyword heuristic to identify cookie banners and click the accept button. Since many breakage incidents happen due to a broken cookie banner, we deemed this interaction necessary. This action does not raise any ethical issues as the website expects this action from any user.

Second, to collect the DOM data, we inject a JavaScript script that reads and parses the DOM tree without triggering any request to the web server. We also use standard OpenWPM data collection methods, which strictly read data from the webpage. In addition, the data we collect is from publicly accessible pages with a fresh browser session (we leave a 20-second buffer between sessions to prevent overload). Hence, there are no possible sensitive data leaks in this step.

Third, while performing interactions, to ensure our interactions don't negatively impact visited web pages (\eg submitting form data), we limit our interactions to one target per sequence.  We also prevent elements from the same salient group from being selected successively and choose the target-interaction pairs at random, weighted by their saliency.
These interaction limitations prevent unintended inputs sent to the server like form filling which requires at least two actions on two distinct targets (fill and submit). 
We recognize that when filling inputs, data may be collected by the server, either as part of auto-completion features (which do not result in storage of unintended information) or exfiltrated as part of tracking collection practices happening before form submission (which we do not consider as an ethical issue as this information should not have been collected in the first place).

\subsection{Differential subtree creation}
\label{subsec:classification}

We process the data we obtain from the crawler visits to create \textit{differential subtrees} as follows:

\para{DOM augmentation.} The output of our interactive visit includes HTTP requests, scripts, DOM trees, and interactions that occurred during a page visit.
We augment the DOM trees with HTTP requests by creating edges between the requests and the requesting element.
For example, if an image element triggers a request to \texttt{image.com}, we add the edge: \texttt{<img>} $\rightarrow$ \texttt{image.com}.

We also add interactions to nodes in the subtree. In the case of errors, we also add the associated error type.
For example, if clicking a button led to a reference error, we add the edges: \texttt{<button>} $\rightarrow$ \texttt{<click>} $\rightarrow$ \texttt{<Reference error>}. 

Given a visit to webpage $A$, we produce an augmented DOM tree $T_A$ and an \textit{environment} graph $G_A$. The graph $G_A$ contains nodes and edges representing more relationships between scripts, requests, and the DOM. One important relationship captured by the graph is whether a script ``touches'' a DOM node, \ie whether the script queried the node at some point during its lifetime. The intuition behind capturing this relationship is that if a script tried to query a DOM node, and then it was blocked or altered, the queried DOM node might be part of the breakage associated with the alteration of the script. Many features from $G_A$ turn out to be important for classification, see Section~\ref{sec:evaluation}.

\para{Subtree extraction.}
Smith \etal~\cite{smith2022blocked} classify whole pages as broken or not. Instead, we opt for classifying sections of a page, referred to as \textit{subtrees}.
This enables us to separate legitimate blocking of ads from actual breakage of functionality within a broken page.
We evaluate the utility impact of the \textit{subtree} approach in Section~\ref{sec:evaluation}.

\begin{figure}[!t]
	\centering
	\includegraphics[width=\linewidth]{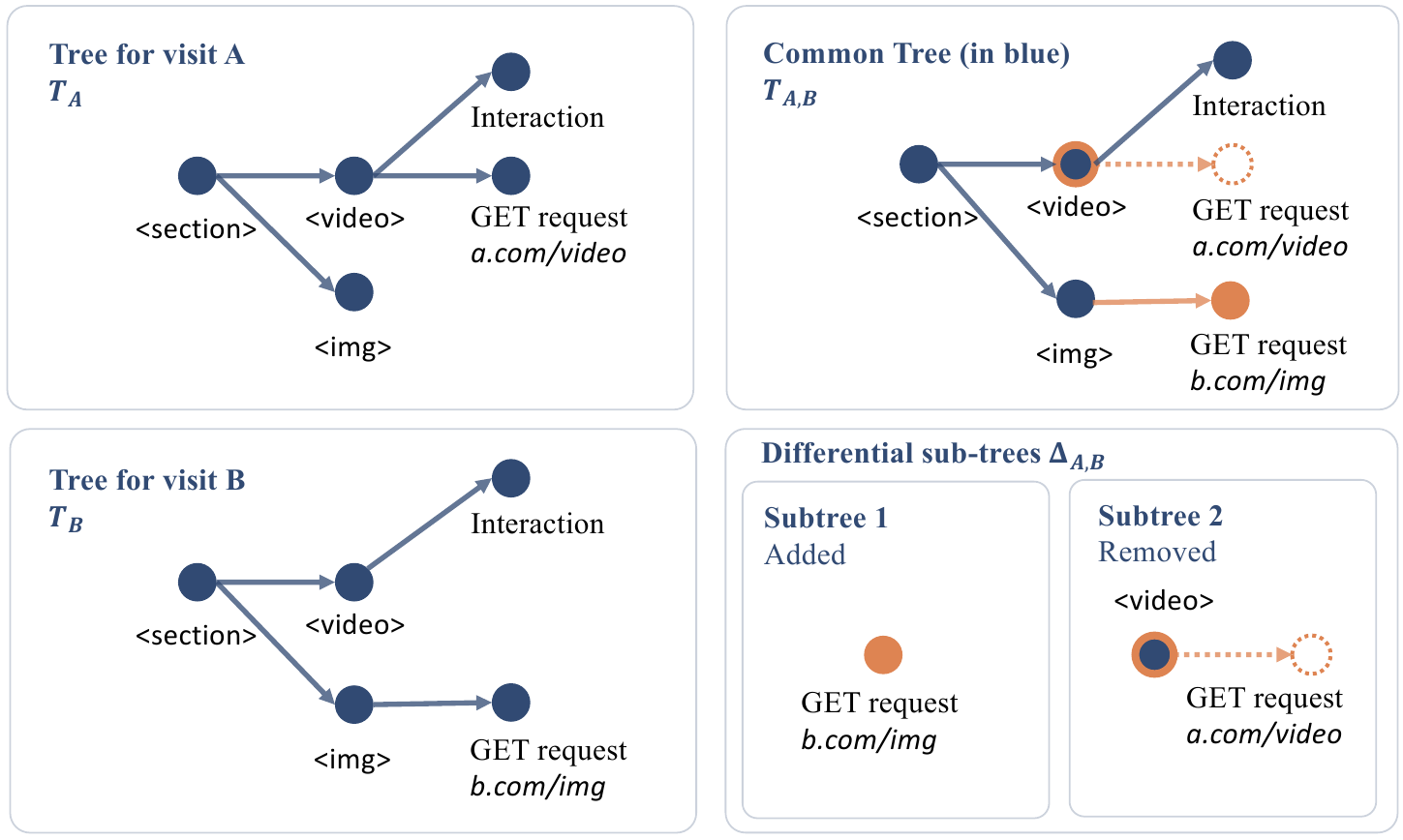}
	\caption[]{Subtree extraction example for the difference going from visit $A$ (top left) to visit $B$ (bottom left). We can see the common tree in blue (top right) $T_{A,B}$ and the differential subtrees set $\Delta_{A,B}$ (bottom right).}
    \label{fig:subtree-extraction}
\end{figure}

Given two DOM trees $T_A$ and $T_B$ for two webpage visits $A$ and $B$, a differential analysis of $T_A$ and $T_B$ returns a common tree $T_{A,B}$ containing all nodes that are the same across the visits, and a set of differential subtrees $\delta \in \Delta_{A,B}$ that represent changes in the tree structure when going from $A$ to $B$. We provide an example in Figure~\ref{fig:subtree-extraction}.
A differential subtree $\delta$ can be of three types: \textbf{ADDED} to $T_{A}$, \textbf{REMOVED} from $T_{A}$, \textbf{EDITED} between $T_{A}$ and $T_{B}$.

\subsection{Subtree classification}
\label{subsec:subtree-classification}
In order to train a classifier that predicts whether a subtree is broken, we extract features from the subtrees and label them. 

\para{Labelling.} 
%
We examine three transitions, $C_{N}$ $\rightarrow$ $C_{F}$, $C_{N}$ $\rightarrow$ $C_{B}$, $C_{F}$ $\rightarrow$ $C_{B}$. 
We label subtrees with one of three possible labels -- \broken, \legit, \neutral -- as follows:

\smallskip\noindent \legit: A \legit label indicates that a subtree was modified by a filter-list rule for legitimate reasons, \ie it was involved in ads or tracking.  Such subtrees are likely to represent an ad/tracker that was blocked/hidden only by changes in the new filter list. 
    We assign this label if a subtree is removed or edited in the visit transitions $C_{N}$ $\rightarrow$ $C_{F}$ or $C_{B}$ $\rightarrow$ $C_{F}$. 
    If a subtree is removed/edited from no filter lists to the breaking filter list ($C_{N}$ $\rightarrow$ $C_{B}$), the label of the subtree is inconclusive, as both broken or legitimately blocked subtrees can undergo this modification from introducing a breaking filter list.
    To have a conclusive label, we need to look at whether the subtree was removed/edited in $C_{N}$ $\rightarrow$ $C_{F}$; if it is blocked in both versions, then we assign the \legit label. 
    
\smallskip\noindent \neutral: A \neutral label indicates that a subtree was modified, but the modification was caused due to factors independent of the filter list, \eg page dynamism. %
    We assign this label if a subtree is added in $C_{N}$ $\rightarrow$ $C_{F}$ or $C_{N}$ $\rightarrow$ $C_{B}$. 
    %
    %
    As rules remove/edit elements, this must be caused by events outside of a filter list. For $C_{B}$ $\rightarrow$ $C_{F}$, the result is inconclusive, as the subtree might also be caused by wrongly-blocked content that got fixed in the new filter list. In this case, we also look at whether the subtree was added in $C_{N}$ $\rightarrow$ $C_{F}$; if it was, we assign the \neutral label.
    
\smallskip\noindent \broken: A \broken label indicates that the subtree represents web elements that were broken due to a filter-list rule. We assign this label if a subtree is removed or edited in $C_{N}$ $\rightarrow$ $C_{B}$ but not in $C_{N}$ $\rightarrow$ $C_{F}$. This implies that there was breakage, as the subtree only exhibited removal/edits on the introduction of the breaking filter list. For $C_{B}$ $\rightarrow$ $C_{F}$, if a subtree is added, the result is inconclusive as it might also be wrongly blocked content that got fixed in the new filter list. In this case, we confirm that the subtree was not added in $C_{N}$ $\rightarrow$ $C_{F}$ to assign the \broken label. If the subtree was added, then we assign the \neutral label.

\begin{figure}[!t]
	\centering
	\includegraphics[width=\linewidth]{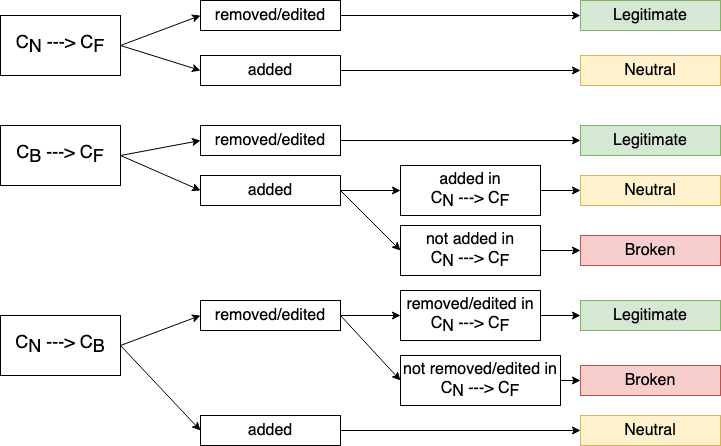}
	\caption[]{Decision tree to label the subtree given the ground truth origin of the visits ($C_F$: visit with fixing filter list, $C_B$: visit with breaking filter list, and $C_N$: visit with no filter lists). The labeling also depends on what happened to the subtree between the two visits in question (\textbf{ADDED}, \textbf{REMOVED} or \textbf{EDITED})}
    \label{fig:subtree-labeling}
\end{figure}

We illustrate the relation of the labels with the visit transitions and subtree behavior in Figure~\ref{fig:subtree-labeling}.

\para{Feature extraction.}
We extract features primarily from the generated subtrees. However, we also use the additional edge information from the \textit{environmental} graph (see Section \ref{subsec:classification}) to extract how scripts, JS errors, and interactions are related between nodes in the subtree and the rest of the page.

We extract features within two \textbf{scopes}: \textit{global} features that are computed over all the subtrees for the same web visit or global statistics from the \textit{envrionment graph statistics (\eg script-node edges)}, and \textit{subtree} features, that are computed per subtree. 
Most \textit{global} features are aggregations of the \textit{subtree} features for all subtrees on the same page. 

We extract four \textbf{categories} of features as follows:

\noindent \textit{1. Content} features, which relate to the content of a node in the subtree.
We divide these features into four groups based on the role of the HTML tags: Layout, Text, Input/Output, and Others. 
Layout tags are related to organizational components like \texttt{<div>} and \texttt{<ul>}. 
Text tags are tags representing any verbal content, like \texttt{<p>} and \texttt{<h1>}. 
Input/Output tags represent either input fields like \texttt{<input>} or information display like \texttt{<video>} and \texttt{<img>}.
For each group, we count the number of nodes in visit $A$, and the number of tags removed, added, and edited from $A$ to $B$. 

\noindent\textit{2. Structural} features are those related to the position of a node within a subtree, a subtree within a page, or connectivity and ancestry relationships among nodes. 
These include features such as the depth of a subtree, the average number and variance of children per node, and the total number of subtrees added, removed, or edited from $A$ to $B$.

\noindent\textit{3. Visual} features that cover changes in subtree size and position on the screen.
We also count the number of salient nodes within a subtree, the number of salient nodes covered by the subtree's footprint on the screen, as well as the changes in these numbers from $A$ to $B$.

\noindent\textit{4. Functional} features are those that relate to crawler interactions and the resulting JavaScript events in a subtree. 
These features capture the effect of script activity and crawler interactions on breakage.
They include errors generated by interactions in a subtree or changes caused by scripts on a page (such as elements created or removed by scripts) computed across the environment graph and the subtrees. 

\para{Classification.}
We classify subtrees as \broken, \legit, or \neutral. 
We experiment with four classifiers: XGBoost, Random forest, SVM (Support vector machines), and MLP (Multi-layer perceptron).

In order to obtain page-level classification from subtree classification, we use two heuristics.
The first heuristic, \name-K$k$, labels a page as broken if we find more than $k$ broken subtrees. 
The second heuristic, \name-R$r$, labels a page as broken if the ratio of broken subtrees to all subtrees is more than $r \over 100$. 



\section{Evaluation}
\label{sec:evaluation}
\subsection{Classification performance}
\label{subsec:eval-performance}

\begin{table*}[t]
\centering
\caption{Validation on \ublock and \easylist datasets. Label: B=\broken, L=\legit}
\label{tab:external-datasets}
\begin{tabular}{lcccccccc}
\toprule
 Evaluation Dataset &   Training Dataset & Reproducible &  AUC &  Accuracy &  B Precision &  B Recall & L Precision &  L Recall \\
\midrule
\easylist &     \adguard &            - & 0.74 &      0.74 &           0.62 &        0.74 &           0.72 &        0.60 \\
\easylist &     \adguard & $\checkmark$ & \textbf{0.80} &      \textbf{0.80} &           \textbf{0.76} &        0.71 &           0.70 &        \textbf{0.76} \\
\easylist &     \adguard &     $\times$ & 0.71 &      0.71 &           0.59 &        0.75 &           0.72 &        0.55 \\
\easylist & \easylist & $\checkmark$ & 0.72 &      0.72 &           0.65 &        0.63 &           0.62 &        0.64 \\
\midrule
\ublock &     \adguard &            - & 0.84 &      0.84 &           0.85 &        0.77 &           0.77 &        0.85 \\
  \ublock &     \adguard & $\checkmark$ & \textbf{0.87} &      \textbf{0.87} &           \textbf{0.95} &        \textbf{0.77} &           0.75 &        \textbf{0.95} \\
  \ublock &    \adguard &     $\times$ & 0.83 &      0.83 &           0.79 &        0.76 &           0.78 &        0.80 \\
  \ublock & \ublock & $\checkmark$ & 0.80 &      0.80 &           0.87 &        0.70 &           0.40 &        0.57 \\
\bottomrule
\bottomrule
\end{tabular}
\end{table*}

\para{Subtree-level evaluation.} 
After labeling the ground truth on the \adguard forum dataset according to Section~\ref{subsec:subtree-classification}, we end up with 3,752 (55\%) legitimate subtrees, 1,712 (26\%) broken edit subtrees, and 1,301(19\%) neutral subtrees. 
To address the imbalance across classes, we try the following re-sampling techniques to augment the training data: \textit{Random Over-sampling} which oversamples, at random, the minority (broken) class; \textit{Random Under-sampling} which undersamples, at random, the majority (legitimate) class; and, \textit{SMOTE}~\cite{chawla2002SMOTESyntheticMinority}, which generates synthetic data points from the minority class as a linear combination of chosen samples.

In conjunction with the different re-sampling techniques, we evaluate multiple classifiers to label trees:
a random forest classifier, 
XGBoost -- known for its robust performance in unbalanced datasets ~\cite{chen2016XGBoostScalableTreea}, 
a support vector machine (SVM) classifier, 
and a basic Multi-Layer Perceptron with 3 layers and 100 nodes per layer. 
We follow standard practices and remove features with constant values or negligible variance, imputing empty values with 0 -- as the features are primarily counts, and applying standard scaling ($x' = \frac{x - \mu_x}{\sigma_x}$). 
For each model, we do a 5-fold cross-validation. 

We find that SMOTE resampling gives the best results and that all classifiers perform similarly (all results are within the standard deviation). We use XGBoost in the rest of our experiments which has an AUC of 86\% $\pm 0.02$, 75\% precision, and 63\% recall on Broken subtrees. 

\begin{figure}[!t]
	\centering
	\includegraphics[width=\linewidth]{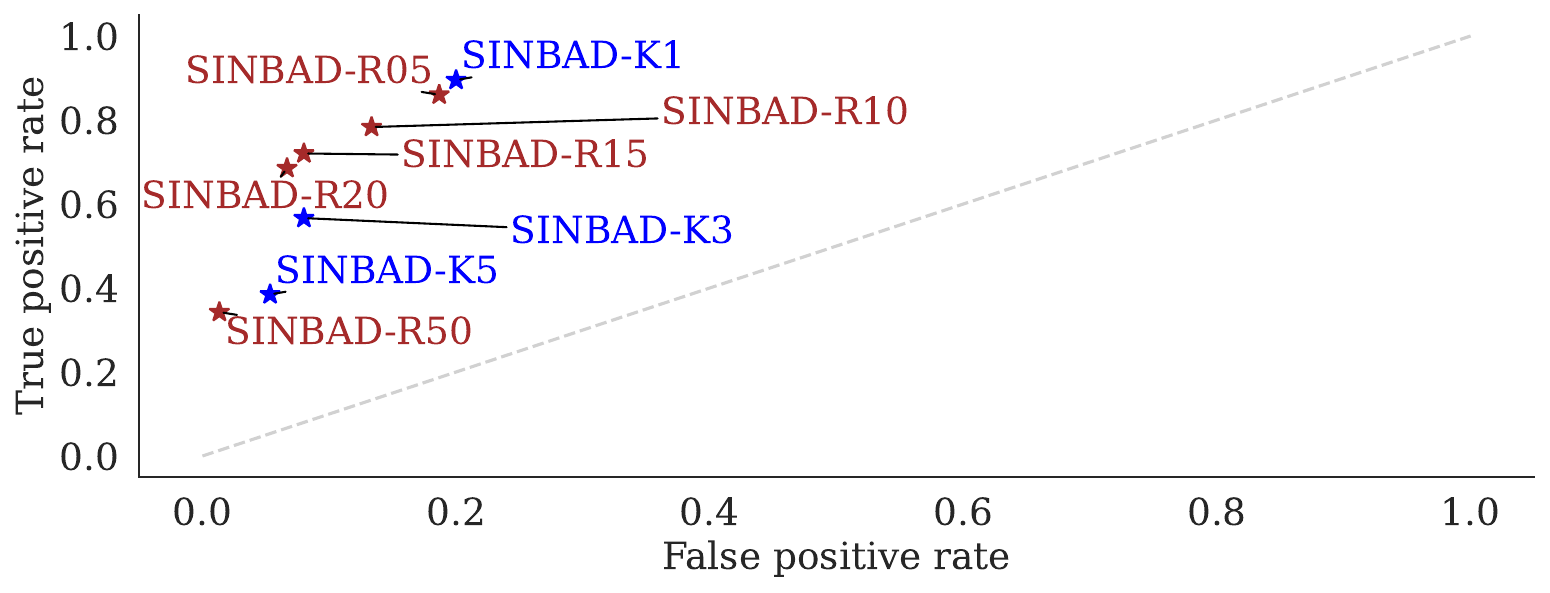}
        \caption{True positive rate and false positive rate of \name page-level heuristics evaluated on the \adguard validation dataset. SINBAD-K$k$ is a count-based heuristic that labels a page as broken if it contains at least $k$ broken subtrees. SINBAD-R$r$ is a ratio-based heuristic that labels a page as broken if the ratio of broken subtrees to all subtrees is at least $r\%$.}
        \label{fig:sinbad-heuristics-performance}
\end{figure}

\para{Page-level evaluation.}
To compare with prior work ~\cite{smith2022blocked}, we perform a page-level evaluation using the heuristics described in Section~\ref{subsec:classification} to  convert tree predictions into page-level labeling of the filter-list change as either breaking or legitimate. We predict breakage on a holdout set of 218 forum issues (143 breaking issues and 75 non-breaking issues). This holdout dataset ensures there is no data leakage using issues that we use to train \name.
We parametrize our count-based heuristic \name-K$k$ with $k \in \{1, 3, 5\}$; the ratio-based heuristic \name-R$r$ with  $r \in \{5, 10, 15, 20, 50\}$. The best-performing heuristic is \name-K1 with an 86\% accuracy and 85\% AUC score (see  Figure~\ref{fig:sinbad-heuristics-performance}). 
Our heuristics perform better with a small value of the parameters because broken subtrees are a minority, and finding at least one is a strong sign of breakage. Increasing the threshold only decreases true positives without having much impact on false positives. 
It is possible that fine-tuned heuristics, more complex crawling implementations, and better-quality datasets would yield even better performance.

\para{Generalization. }
We validate our results by evaluating \name on \easylist and \ublock. We train models on these datasets, and also test whether the model trained on \adguard transfers well. We train both on reproducible issues and on all issues. We report the results in Table~\ref{tab:external-datasets}.

Overall, \name generalizes well. 
On \easylist, the model trained on \adguard achieves an 80\% AUC on reproducible issues, while it drops to 74\% for all issues. Retraining on \easylist results in a performance decrease, which we attribute to the low number of broken subtrees in \easylist (992 broken sub-trees). 
On \ublock, the model trained on \adguard achieves an 87\% AUC and a high 95\% broken subtree precision on reproducible issues, while it drops to 84\% on all issues. Re-training the model on \ublock, on the other hand, results in 80\%, where the decrease stems again from the reduced set of training data (only 783 broken subtrees in \ublock).

\subsection{Feature Analysis}

We analyze the feature importance of XGBoost, ranking features based on the AUC loss metric. 
We report the top-10 ranking in Table~\ref{table:sinbad-feature-importance-small} (full ranking in Table~\ref{table:sinbad-feature-importance} in Appendix).

\begin{table}[b]
\centering
\caption{Top features according to AUC Loss predictive power. \\Scope: S=Subtree, G=Global. Category: V=Visual, S=Structural, F=Functional, C=Content}
\label{table:sinbad-feature-importance-small}
\begin{tabular}{lllll}
\toprule
 &  Scope &    Cat. &                                                                                      Description \\
\midrule
1  &   S &      V &  Subtree page coverage before the filter rule edit \\
2  &   S &      V & Number of salient elements in the subtree \\
3  &   S &  S &  Average degree of subtree nodes \\
4  &  G &  F &  Number of requests added \\
5  &   S &  F & $\Delta$ in elements queried by a sub-tree-related script. \\
6  &  S &  F & $\Delta$ in elements queried after subtree interactions. \\
7  &   G &  F & Number of requests removed t \\
8  &   S &  C & Number of tags considered \textit{Text} added \\
9  &  S &  F &  Number of interactions with the subtree. \\
10  &  S &    C & Number of tags considered \textit{Text} removed \\
\bottomrule
\end{tabular}
\end{table}

The top features of the classifier do not belong to a specific feature category or scope.
Their diversity
indicates that classifying breakage is multifaceted, \ie detecting breakage cannot be reduced to one aspect of the page visit, \eg only \textit{visual} features. 
We highlight that the ``number of salient elements" ranks $2^{\text{nd}}$, confirming that saliency is used to prioritize breakage detection. 
We provide further evidence of the importance of saliency in Appendix~\ref{subsubsec:dynamic-breakage-example}.
We also see that features generated from interaction on the DOM are very relevant, exemplified by the number of changes in the number of elements queried by scripts after an interaction (ranked \eg $6^{\text{th}}$.
This validates our intuition that capturing these relationships, via our DOM tree augmentation, is helpful to detect dynamic breakage.
Finally, counts of elements with visual impact on the user (\eg text modifiers, elements that determine the layout, or forms and figures) have a strong AUC loss impact. We conjecture it is because they capture the semantic role of the subtree within the page. When those user-relevant elements are removed, they trigger breakage.

\subsection{Manual analysis of errors}
\label{subsec:error-analysis}
\begin{figure}
    \centering
    \includegraphics[width=\linewidth]{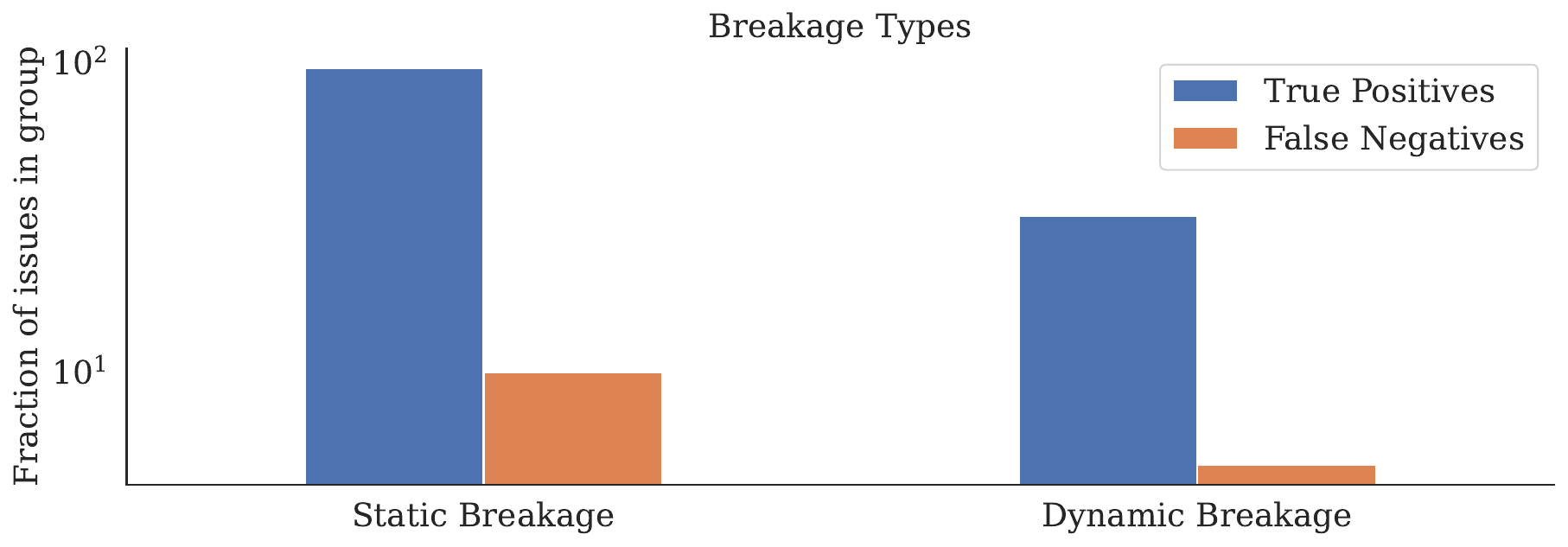}
    \includegraphics[width=\linewidth]{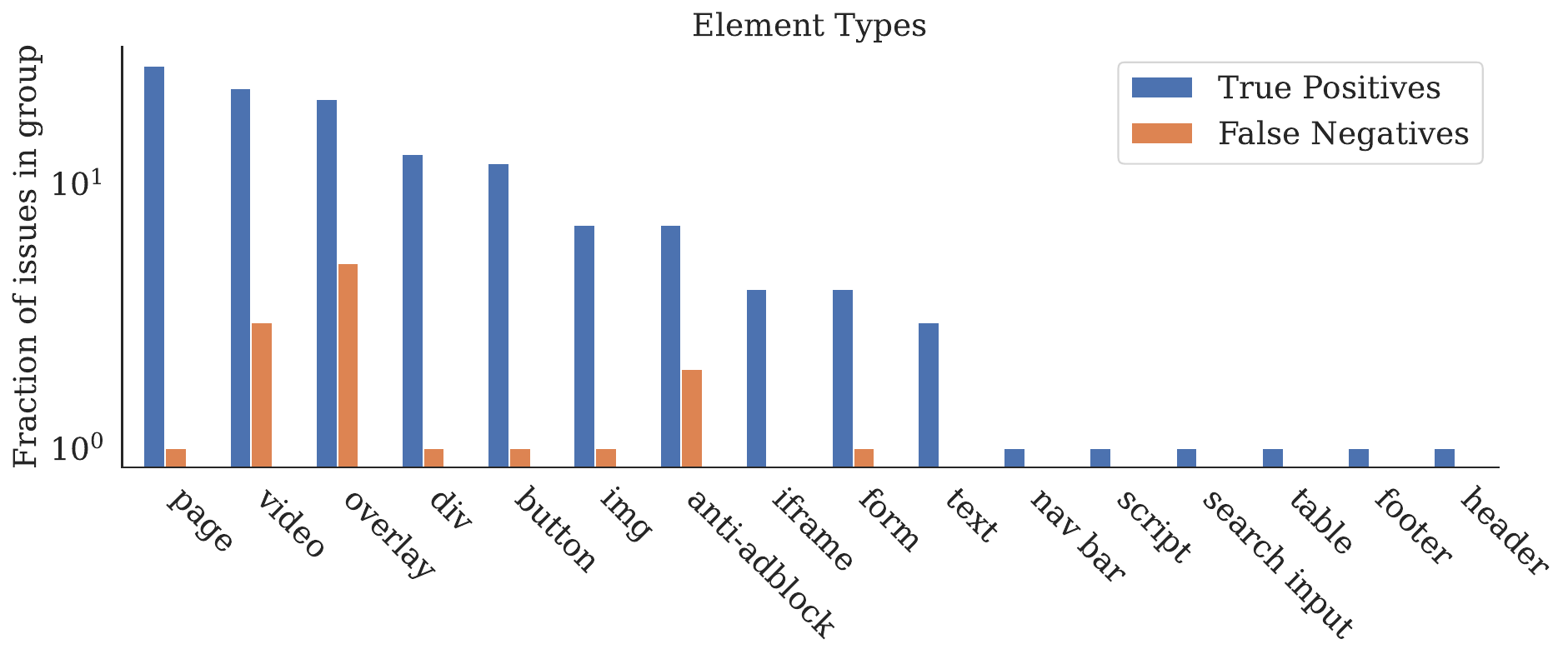}
    \caption{SINBAD errors depending on breakage type (top) and type of element broken (bottom)}
    \label{fig:issue-error-stats}
\end{figure}

We now investigate the errors made by \name and \name-K1 on the validation forum issues in Section~\ref{subsec:eval-performance}. 

\para{False negatives (missed breakage issues).}
We investigate the 15 breakage issues that \name did not detect. We acknowledge that 15 false negatives is a small number and can affect the significance of our statistical claims, but these negatives are a 6\% of the 218 issues we use for validation. 

We plot in Figure~\ref{fig:issue-error-stats} the distribution of false negatives across breakage types (static/dynamic) and across the type of broken elements we manually labeled. 
Only 9\% of static breakage (10 issues) and 13\% of dynamic breakage (5 issues) went undetected. 
These errors are due to two reasons. First, \name could not perfectly reconstruct the filter list in five of these issues. For example, in issues involving the use of unknown user-defined rules, or involving filter lists that we could not track back in time. 
Second, seven of the issues are not reproducible (despite reproducibility checks in Section~\ref{subsubsec:reproduce-analysis}) and thus could not be identified. 
This was due to page updates, geographic locks, pages disappearing, or domain hopping. 
We discuss in Appendix~\ref{domain-hopping-example} how this practice brings many complications for breakage detection.

We find that the three remaining false negatives are edge cases. 
The first issue was due to an anomaly in the maintainer's workflow in which a commit was reverted and thus the last commit in the forum did not correspond to the fixing commit. Thus, \name was testing breakage with the wrong filter list. 
In the second issue, a website still was capable of using anti-adblock despite fixes in the filter list causing anomalous behavior. 
As we could classify correctly eight issues with anti-adblock (Figure~\ref{fig:issue-error-stats}), we assumed this is an outlier and did not delve deeper into the problem. 
The third issue is caused by the presence of \textit{overlays} -- full-page backgrounds that are usually gray or transparent.
Users experience breakage if they cannot close an overlay and access the main page content. 
An overlay can be wrongly included by \name as essential content in edge cases, rather than designating it as the cookie banner and closing it, due to language-related limitations (\eg, buttons in non-English language, more details in Appendix~\ref{subsubsec:overlay-breakage-example}).

\begin{figure}
    \centering
    \includegraphics[width=\linewidth]{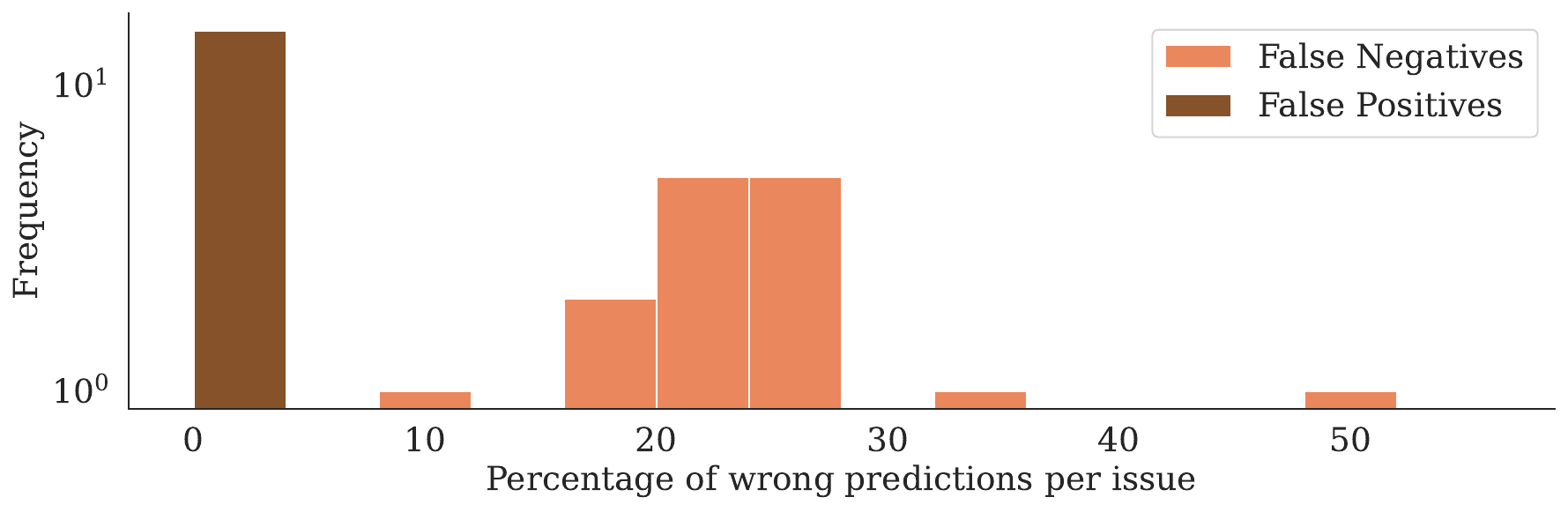}
    \caption{Distribution of the percentage of wrongly predicted sub-trees by \name from all sub-trees in issues for both false positives and false negative issues by \name-K1}
    \label{fig:wrong-subtree-percentage}
\end{figure}

\para{False positives (issues misclassified as broken). }
\name-K1's false positives are harder to group into specific categories. 
This method has a strictness imbalance between classifying something as breakage and not: misclassifying a page as broken requires classifying just one subtree as broken, while misclassifying it as not broken requires (potentially) many non-broken predictions. 
Thus, it takes more than 10\% prediction errors by \name to misclassify a page as legitimate, and only 4\% of errors or less to cause a false breakage alert (see Figure~\ref{fig:wrong-subtree-percentage}). 
More complex heuristics could lead to more balanced errors.  

In our experiments, most errors come from earlier parts of the pipeline, including filter-list reconstruction and crawling, rather than classification. 
Five issues involved requests to remove empty ad containers left behind from correctly blocked banner ads. 
Such issues illustrate a vague boundary between breakage issues and ad-blocking requests.
Some users may consider this as breakage, whereas others may simply view it as a partially blocked ad. 
We also find errors on two sites due to the presence of a large amount of randomly changing content across visits, \eg image gallery with shuffled images.
This can result in the creation of many subtrees (even more than 40) which increases the chances that \name mislabels as broken.

\begin{figure}
    \centering
    \includegraphics[width=\linewidth]{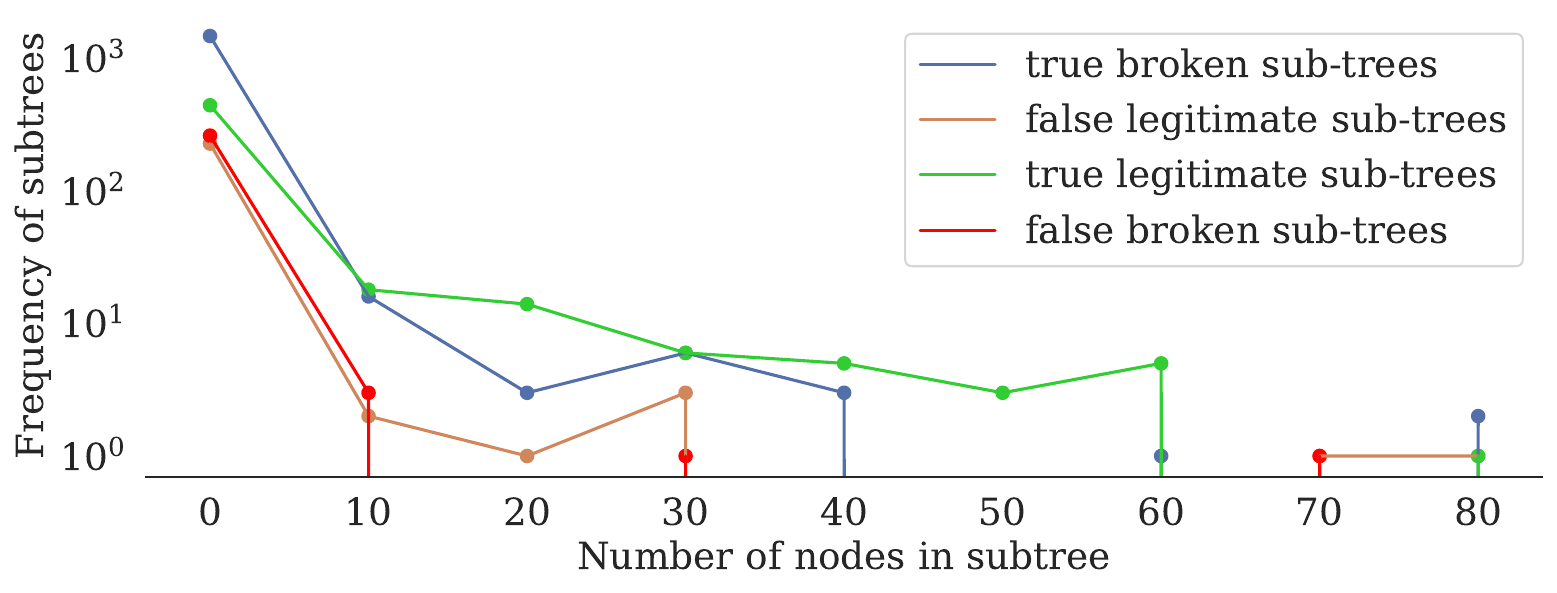}
    \caption{Distribution of the number of nodes per subtree across four groups of \name predictions: true broken subtrees, false legitimate subtrees, true legitimate subtrees, and false broken subtrees.}
    \label{fig:n_nodes-distribution}
\end{figure}

\para{Subtree misclassifications.}
We examine the distribution of features in the misclassified subtrees where errors did not originate from crawling or issues in filter-list recovery from forum posts. 
We observe that these subtrees have few nodes (see Figure~\ref{fig:n_nodes-distribution}), and thus provide little information, hindering classification.
In fact, these small subtrees are usually wrongly labelled by our crawler as \texttt{<svg/>}, creating noisy ground truth that induces errors. 
This is a limitation of our node-similarity heuristic, which we discuss further in Appendix~\ref{subsubsec:similarity-heuristic}.

\subsection{Comparison with existing detectors}
\label{subsec:compare-methods}
In this section, we compare \name against existing methods to detect breakage.
Previous studies quantify breakage automatically via two approaches: (1) using heuristics based on the changes in the page resources loaded (number of images, change in the website text \etc)~\cite{krishnamurthy2007measuring, fouquet2023breaking, le2022autofr}, network traffic differences~\cite{amjad2023blocking}, or visibility changes in webpages~\cite{castell2023astrack}; and (2) machine-learning-based approach using a graph-based representation of webpages~\cite{smith2022blocked}. 
Unlike \name, none of these approaches can detect cases of dynamic breakage.
We compare \name against these approaches on the \adguard validation set. 
As previous methods operate on at page level, we use \name's best site-based heuristic, \name-K1. 

\begin{figure}[!t]
	\centering
	\includegraphics[width=\linewidth]{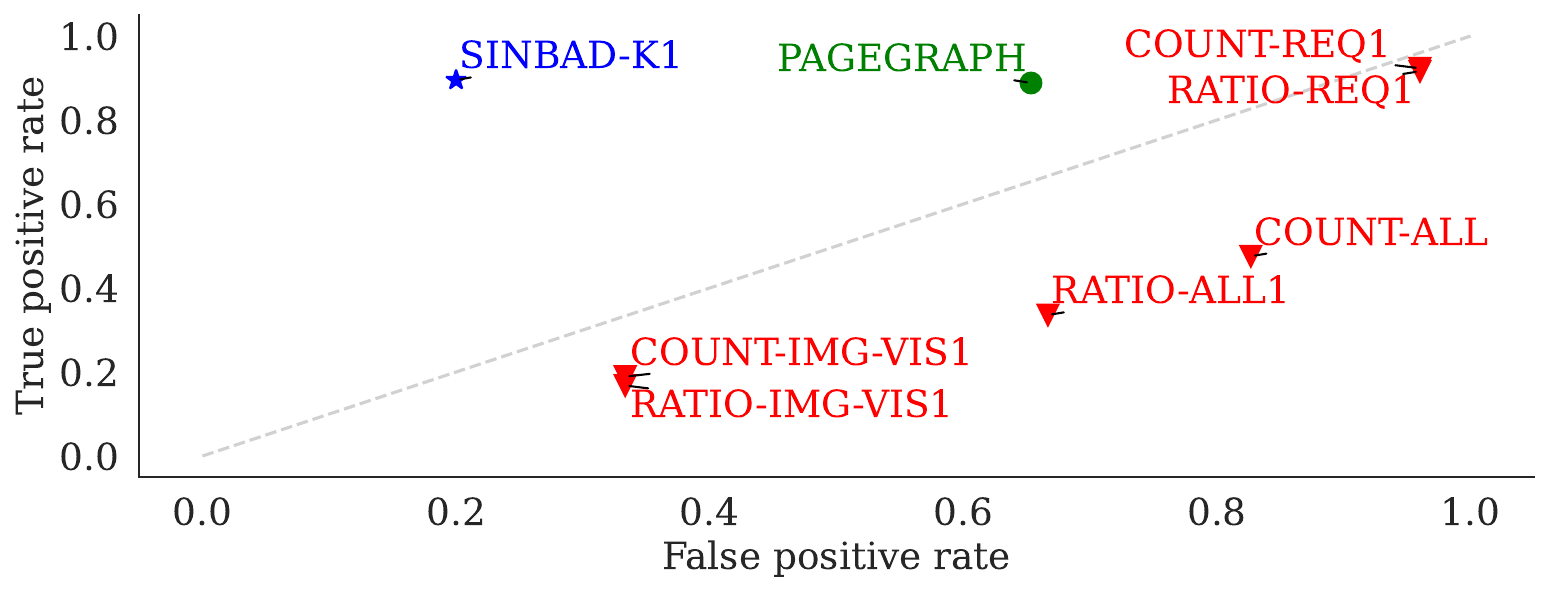}
        \caption{True positive rate and false positive rate of the main heuristics evaluated on the \adguard validation dataset. (Blue=\name; Green=\texttt{PAGEGRAPH} (Smith \etal~\cite{smith2022blocked}); Red=Heuristic-based approaches)}
        \label{fig:heuristics-performance}
\end{figure}

\para{Heuristics-based approaches.} 
Inspired by previous works, we use two similar classes of heuristics: count-based threshold heuristics~\cite{le2022autofr, amjad2023blocking}:

\begin{small}
\[\texttt{COUNT-<elements>}k: \]
\[
\begin{cases}
\texttt{breaking} & \text{if any } \text{Avg}_{\text{e} \in \text{elements}}(\Delta \ \# \text{e}) > k\\
\texttt{non-breaking} & \text{otherwise} \\
\end{cases}\]
\end{small}

\noindent
and ratio-based threshold heuristics~\cite{krishnamurthy2007measuring}:

\begin{small}
\[\texttt{RATIO-<elements>}r:\] 
\[\begin{cases}
\texttt{breaking} & \text{if any } \text{Avg}_{\text{e} \in \text{elements}}(\frac{\Delta\# \text{e}}{\# \text{all e}}) > \frac{r}{100}\\
\texttt{non-breaking} & \text{otherwise}
\end{cases} \]
\end{small}

The heuristics we evaluate are \texttt{COUNT-REQ1} and \texttt{RATIO-REQ1} (network requests), \texttt{COUNT-IMG-VIS1} and \texttt{RATIO-IMG-VIS1} (visible images), \texttt{COUNT-ALL} and \texttt{RATIO-ALL1} (images, buttons, and text).
%
As shown in Figure~\ref{fig:heuristics-performance}, our model outperforms these heuristics by orders of magnitude. 
We experimented with other heuristic threshold values, but they performed worse than those in the figure. 

No heuristic achieves a result above 55\% accuracy and AUC score, as opposed to \name's 86\% accuracy and 85\% AUC score. 
All of them fail to detect legitimate changes on the website, with precision and recall as low as 10\% and 20\%. This is because these classifiers make no difference between blocking an annoying element and an important (salient) element on a page.
%

\para{Machine-learning-based approach.}
We compare \name against Smith \etal\cite{smith2022blocked}, the only machine-learning approach for breakage detection to the best of our knowledge, who report an AUC of 88\%.

We find that Smith \etal's methodology leads to a noisy ground truth for two main reasons, and thus their AUC may not be representative of actual breakage detection.
First, they include issues as old as 2013, but given that their experiments were run in 2022, many of these issues may not have been reproducible at the time of the crawl (our manual reproducibility investigation of \easylist  shows that reproducibility drops quickly with time, down to 27\% after 2 years, see Section~\ref{subsubsec:reproduce-analysis}). 
We also find that Smith \etal.'s heuristic intended to avoid this pitfall -- using variation in the network traffic across two crawls (before and after a filter list change) -- is unreliable.
On 170 \easylist issues, of which 44\% are reproducible and 56\% are not, 55 out of 66 non-reproducible posts are falsely labeled as reproducible by Smith \etal's heuristic. This casts doubt on the representativeness of the ground truth used in \cite{smith2022blocked}.
%
%

The second issue comes from the data curation approach. Smith \etal. directly scrape commits from the filter-list repository, and treat each commit that mentions the issue as a fixing commit. Yet, 18\% of the issues need two or more commits from a maintainer moderator before the issue is resolved (see Section~\ref{subsec:dataset-scraping}). This means that many of the examples used by Smith \etal. may not have been actually fixed.
To address this issue, Smith \etal. filter commits that mention ``fix'' in the title and expect the broken page URL to be mentioned in the commit title. 
%
But, we find that 20\% of 1344 \easylist issues state the base domain of a broken page rather than the full URL in the commit title, which would have led to errors in the ground truth in \cite{smith2022blocked} possibly overestimating the number of broken samples.
%

%

We attempted to test Smith \etal's pipeline on our validation dataset so as to have a head-to-head comparison with \name.
Unfortunately, we were unable to run the crawling and graph creation code due to missing and deprecated dependencies, even after several rounds of communication with the authors. 
Thus, we choose to re-implement the features used in their classifier. 
We succeeded at implementing 32 out of their top 40 features and failed to implement those that are related to their custom webpage representation~\cite{PageGraph}, which we would have to reproduce as well.
We train Smith \etal's model on the same dataset that we train \name model on to minimize the dataset effect on the trained models. We call this model \texttt{PAGEGRAPH}.

On the \adguard validation dataset used throughout Section~\ref{sec:evaluation}, \texttt{PAGEGRAPH} only achieves 65\% accuracy and a 57\% AUC score, compared to \name's 86\% accuracy and 85\% AUC score. 
While some \texttt{PAGEGRAPH}'s performance loss with respect to the AUC reported in the paper can be attributed to the features we could not implement, given that these features had little importance, we believe this difference is not relevant.

To ensure that the advantage of \name is not tailored to our validation dataset, we repeat the comparison using three other datasets: \texttt{ADGUARD-24} (100 newer Adguard issues starting on the 28$^{\text{th}}$ Nov. 2023, 50 breakages), \texttt{EASYLIST} (120 issues, 57 breakages), and \texttt{UBLOCK} (31 issues, 12 breakages). For the latter two, due to the lack of non-broken examples, we considered non-reproducible issues as non-broken which we manually checked is a good approximation. We report the results of the comparison in  Table~\ref{table:new-datasets-compare}, where wer see that \name outperforms \texttt{PAGEGRAPH} for all three datasets. We conclude that \name's advantage over \texttt{PAGEGRAPH} is dataset-independent. 

\begin{table}[t]
\centering
\caption{Page-level comparison between \name and our re-implementation of \texttt{PAGEGRAPH} by Smith \etal~\cite{smith2022blocked}}
\label{table:new-datasets-compare}
\begin{tabular}{lllll}
\toprule
 Dataset &  Model &   AUC & TPR & FPR \\
\midrule
\texttt{ADGUARD}    & \name & \textbf{0.85} & 0.89 & 0.20\\
                    & \texttt{PAGEGRAPH} & 0.57 & 0.88 & 0.61\\
\midrule
\texttt{ADGUARD-24} & \name & \textbf{0.87} & 0.94 & 0.20\\
                    & \texttt{PAGEGRAPH} & 0.55 & 0.86 & 0.76\\
\midrule
\texttt{EASYLIST}   & \name & \textbf{0.78} & 0.82 & 0.25\\
                    & \texttt{PAGEGRAPH}& 0.48 & 0.89 & 0.94\\
\midrule
\texttt{UBLOCK}     & \name & \textbf{0.88} & 1.00 & 0.24\\
                    & \texttt{PAGEGRAPH}& 0.39 & 0.75 & 0.90\\
\bottomrule
\end{tabular}
\end{table}

\subsection{\name in an open-world setting}
\label{subsec:open-world}
We evaluate \name in an open-world setting to assess its ability to reliably find breakage in the wild.
We study the potential effects of two kinds of rules: site-specific rules written by mantainers to avoid creating issues in other sites, and generic rules aimed to affect all sites (\eg blocking generic tracking APIs).

We collect 106 websites (50\% top and 50\% random sites from Alexa's top-1M) over 3 generic and 3 site-specific filter-list changes from the Adguard's forum. Among the 106 sites, We found one unreported breakage for one of the generic settings, 5 to 8 false positives in the site-specific settings, and 7 false positives with a generic filter-list change. We manually inspect these false postives and find that same very popular sites (\eg ebay.com, amazon.com, and imdb.com) are triggering false alarms due to two issues:

First, an implementation issue results in some requests in some visits being recorded twice. When this double counting only happens in one visit, \name interprets that the element has been \textbf{REMOVED} or \textbf{ADDED}, even though the element has not changed in reality. A similar problem happens when webs use SVGs, as they load differently between visits and are labeled \textbf{EDITED} even if they are the same. Removing these falsely created sub-trees reduces the average false positives to 3 (FPR of 2.9\%).

Second, some pages return random content on each reload (\eg youtube.com recommended videos) which appear as \textbf{REMOVED} and \textbf{ADDED} between two visits and causes \name to predict a breakage. This limitation is not \name-specific, but an inherent challenge to the differential approach that affects all existing works, including Smith \etal \cite{smith2022blocked}. It could be solved by reloading the page several times to detect and ignore ever-changing content. If this issue would be resolved by the community, \name's false positive rate would be 0.63\%. 

In summary, \name's open world FPR is much lower than in the controlled experiments, mainly because it is unlikely that a filter-list change impacts random webs and erroneously produces sub-trees. We conclude that \name is a promising approach for maintainers to automate the discovery of breakage and avoid deploying harmful changes.


\subsection{Efficiency}
%
We measure the overhead of \name's large-scale crawls and graph creation processes to understand if is is suitable for deployment during filter-list rule creation.

The dominating overhead in \name is crawl time. 
For our dataset, \name took an average of $53 \pm 35$ seconds for the first crawl (with the fixing filter-list and saliency predictions), $48 \pm 20$ seconds for the second crawl (with the breaking filter list) list, and $43\pm 18$ seconds for the third crawl (with no filter list), per page.
%
The entire crawl, without parallelization, took 23 hours for our dataset of 543 sites, with a large variance from site to site, due to the difference in the number of DOM nodes across sites and in the time to fetch them from the server.
These numbers can be reduced by decreasing the timeout threshold (we use 400 seconds), and by parallelizing the crawl.

Per site, \name takes $6.3 \pm 1.8$ seconds to build the trees, $9.3\pm 9.0$ seconds to extract subtrees, $1.3\pm3.1$ seconds to obtain features,and  $3.7\pm0.9$ milliseconds to test against the classifier. Thus, it would take about 1 day to test a filter list on the top-10k sites, running on 20 instances. This performance is sufficient to keep up with the current update frequency of popular lists. These numbers could be improved by carefully optimizing the code.

\section{Take aways}
\label{sec:discussion}

In this work, we have introduced a new automated tool for detecting breakage, \name. Trained on verified user-reported breakage instances that we extract from blocking tools forums, \name improves significantly over the state of the art, both in terms of accuracy -- with a 20\% increase, and coverage -- detecting dynamic and visual figures that were missed by previous work.

\para{Usage scope.}
\name can be used by filter-list maintainers to check the breakage potential of new rules before deployment.
\name's detection granularity makes it easy for maintainers to identify problematic rules and adjust them.
\name can also be used to augment automated advertisement and tracking detection tools~\cite{iqbal20adgraph, iqbal21fingerprinting, iqbal22khaleesi, siby22webgraph, munir23cookiegraph, yang2022wtagraph} with tests to understand whether removing the identified resource would result in breakage; or automatic filter-list rule creation~\cite{le2022autofr} with tests to detect problematic rules. 

\name is not restricted to breakage caused by ad-blocking filter lists. It can be adapted to any other breakage source, provided that there are good data sources to train the classifier. The source code and instructions for \name can be found at \url{https://github.com/spring-epfl/sinbad}.

\para{Modularity}
\name follows a modular design. Every component can be substituted if a better alternative becomes available. 
For instance, a future saliency model with better performance, or a better segmentation algorithm. \name also has an interface to create custom interactions with minimum restrictions, to augment the crawler with new instructions such as filling login forms~\cite{senolLeakyFormsStudy, roomiLargeScaleMeasurementWebsite}.

\para{Future Improvements.} 
Breakage forums' data are very reliable, but it is hard to obtain a large number of samples, and there is no guarantee that the ones obtained are representative of all possible types of breakage that users face. Integrating a reporting mechanism in blocking tools for users to report breakage in a structured manner can increase the amount of data available and capture more forms of breakage~\cite{nisenoff2023defining}.

Data availability is also affected by reproducibility -- due to natural page dynamism or due to the issue having been fixed before the crawl. More work is needed to find ways to recreate breakage (\eg using a crowd-sourcing tool or web extension for users to report breakage and store visit snapshots with enough details for the features). 

\small
\bibliographystyle{IEEEtran}
\bibliography{IEEEabrv, bibliography}
\normalsize

\appendices

\section{Examples}
\begin{figure*}
    \centering
    \includegraphics[width=0.25\linewidth]{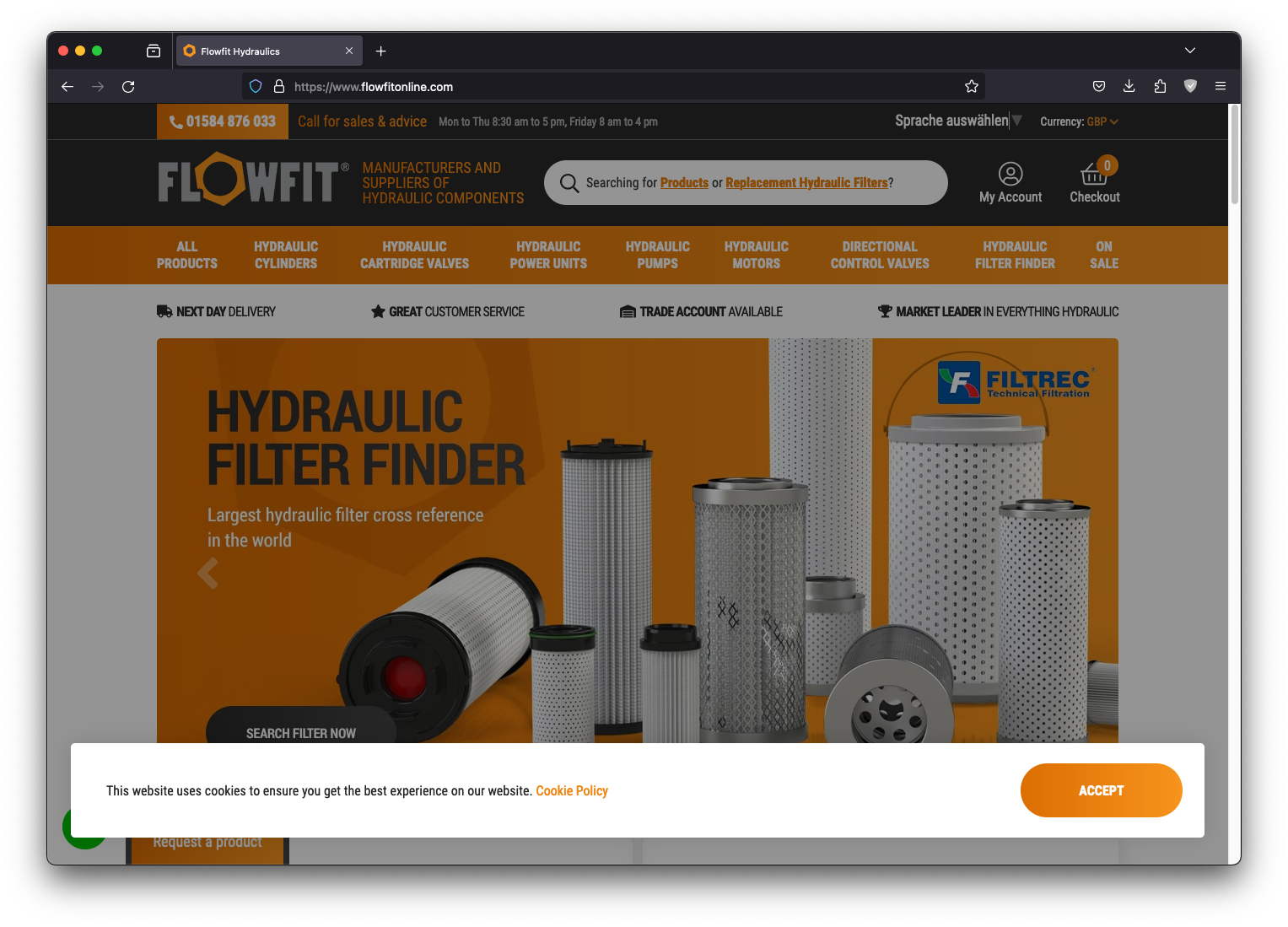}
    \includegraphics[width=0.25\linewidth]{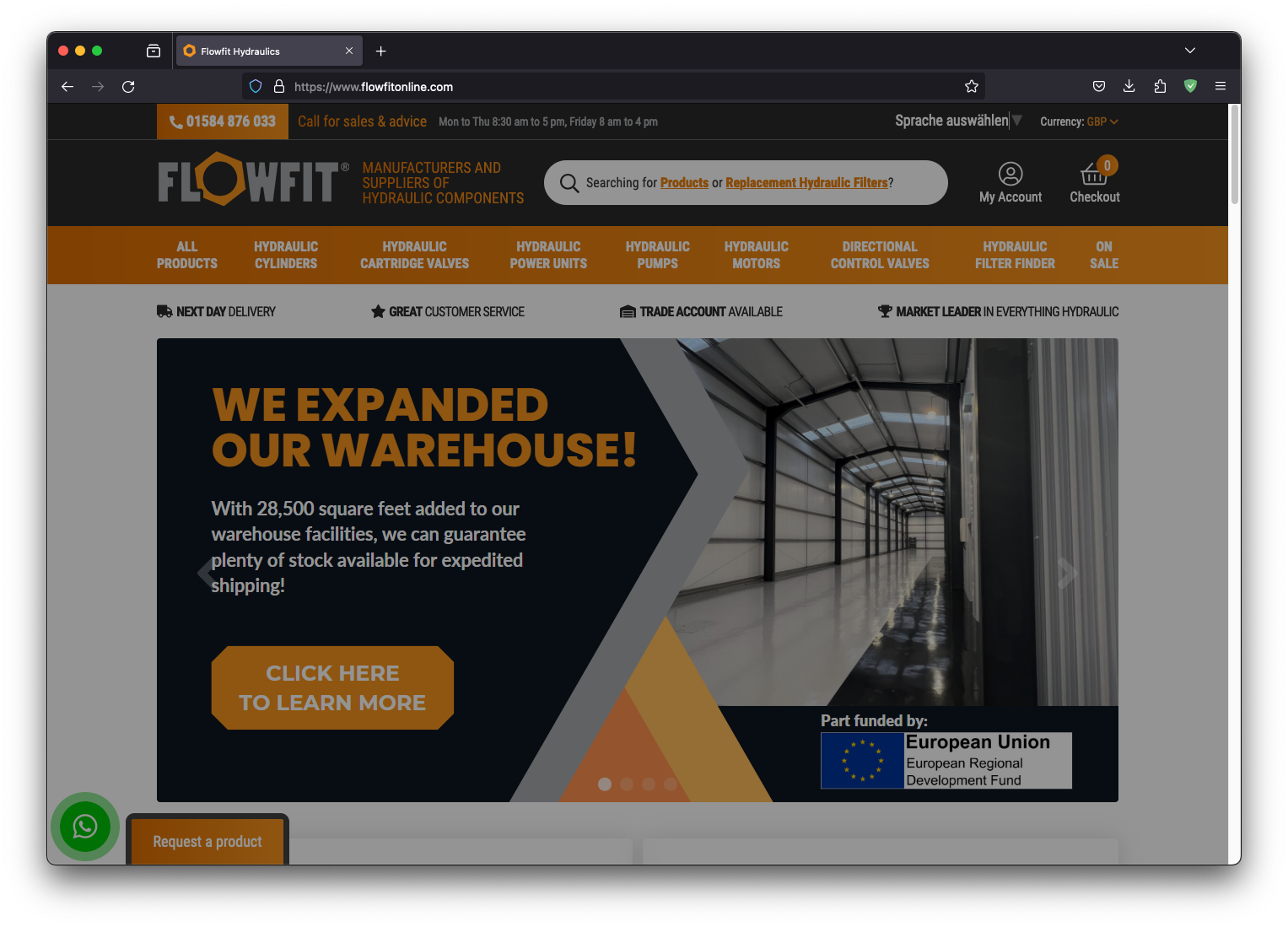}
    \includegraphics[width=0.25\linewidth]{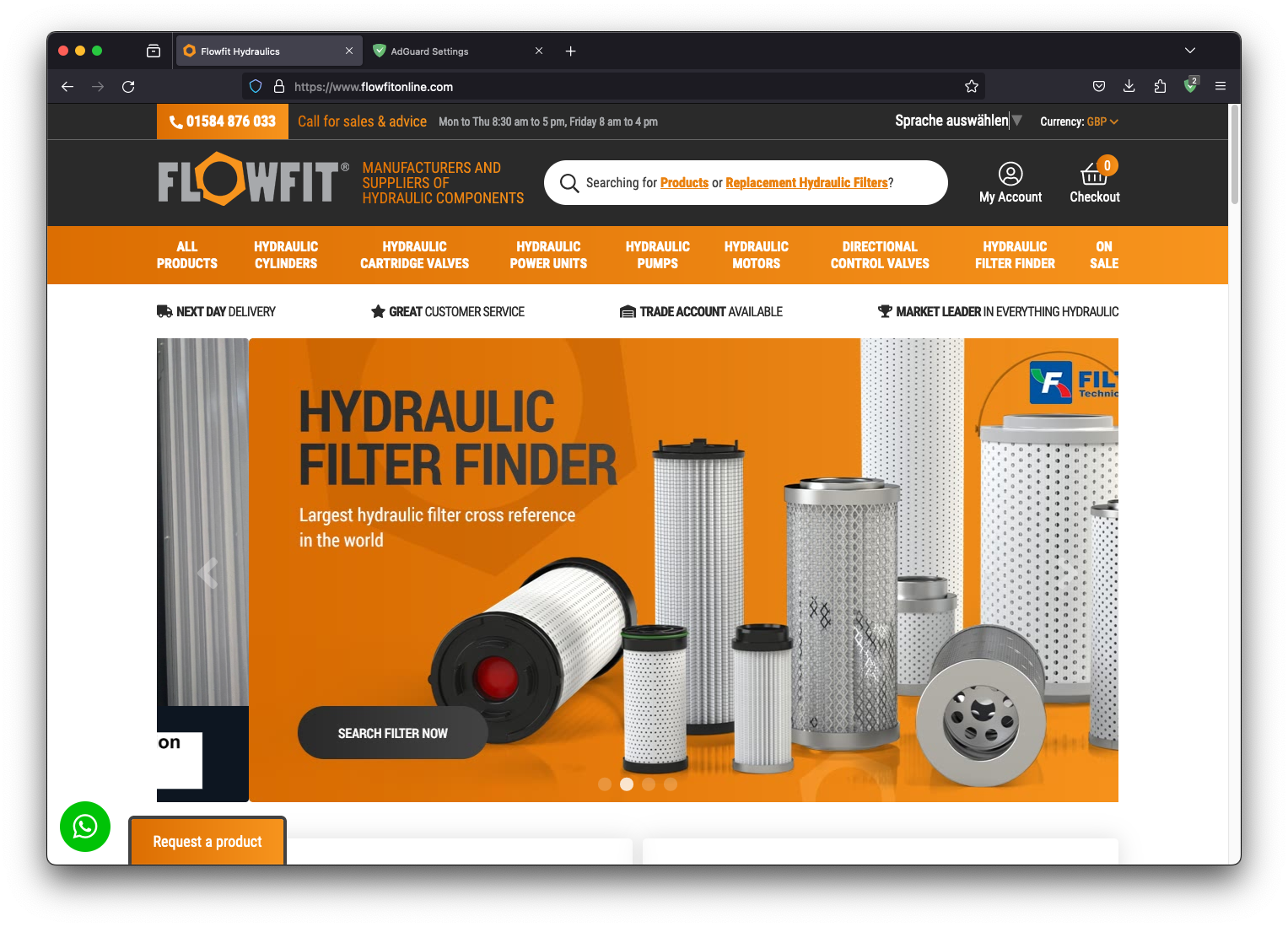}
    \caption{Overlay breakage example. \textbf{Left:} site without any filter rule. \textbf{Middle: } site with broken filter rules. \textbf{Right: } site with the fixing filter rules.}
    \label{fig:example-overlay-breakage}
\end{figure*}
\subsection{Interdependencies in filter lists}\label{sec:interdependency}

We provide a breakage example that is not triggered by the reported breaking rule but requires the whole filter list to be reproduced.
We take the \adguard issue \textit{\#162559}, where the user loads 11 collections of filter lists. The reported breakage is a dark overlay over a phone screen depiction on the website. The fixing commit adds the following rule 

\begin{lstlisting}[caption={Adguard fixing commit change},label={lst:adblockplus},breaklines=true]]
+ app.programme.conventus.de##ion-app > #ion-overlay-1
\end{lstlisting}
This rule blocks any HTML element whose parent has the id ``\texttt{ion-id}'' and has the id ``\texttt{ion-overlay-1}''. 
We experiment with three different filter-list setups using AdGuard on Firefox. For filter list A, we reverse the rule to get the opposite effect as Smith \etal suggest~\cite{smith2022blocked}. 
For filter list B, we only take the edited filter list Adguard's \texttt{cookies\_specific.txt}. 
For filter list C, we include all the filter rules in the configuration at the time of the issue. 
Only the third setup succeeds in reproducing the issue. 
The other setups show a cookie notice where the breakage should have occurred. The reason is that a global filter list installed by the user, left unchanged by the maintainer, already blocked the cookie consent form without blocking the overlay \texttt{ion-overlay-1}. The fix involved hiding the additional overlay which is written into \texttt{cookies\_specific.txt}, and if we only use \texttt{cookies\_specific.txt} on the website (setup B), only the overlay will be blocked and not the cookie consent form. Both the global filter list and the fixed \texttt{cookies\_specific.txt} must be present (setup C) to have both elements blocked. 

\subsection{Dynamic breakage caught due to saliency and interactions}
\label{subsubsec:dynamic-breakage-example}

In \adguard issue \texttt{139618}, \name correctly predicts 39 subtrees our of 46 -- 16 \broken,  7 \legit, and 16 \neutral. Among the broken subtrees, we get \textit{iframes}, \textit{scripts}, and most importantly, an edited \textit{div} which is the parent of a video container that does not load correctly. The forum indicates that the video does not load because it depends on a script that runs as a response to accepting cookies. As the filter rules hide the cookie banner, this script is never triggered, resulting in no video. \name finds this video as a salient element and attempts to click on it. In the fixed version, the video loads correctly. \name captures this difference in behavior (change in elements touched due to an interaction \texttt{dn\_el\_in\_int\_tree}, change in the number of visible elements \texttt{n\_visible\_rem}), and marks this subtree as breaking. The falsely labeled elements are small \texttt{<svg/>} due to reasons described in Appendix~\ref{subsubsec:similarity-heuristic}.

\begin{figure*}
    \centering
    \includegraphics[width=0.7\linewidth]{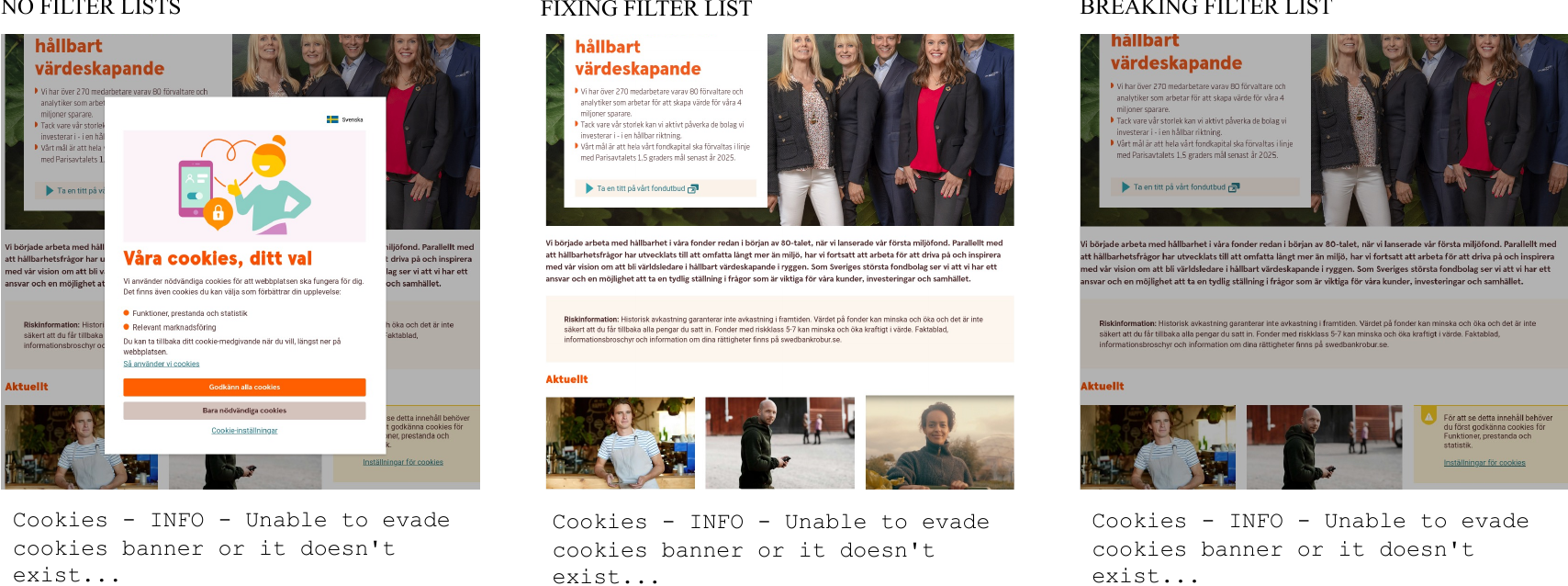}
    \caption{Example of an ``overlay'' breakage undetected by \name-K1. The \textit{Cookie-banner} crawling module failed to detect the cookie banner.}
    \label{fig:overlay-breakage-failed}
\end{figure*}

\subsection{Content-rule Breakage undetected by network-based features}
\label{subsubsec:cosmetic-breakage-example}
In \adguard issue \texttt{157392}, \name correctly predicts all 11 subtrees extracted -- 3 \broken, 3 \legit, and 2 \neutral. The broken subtrees are an edited search bar, a header, and the main content \texttt{div}. The forum indicates that the breakage happens due to an anti-ad-block employed by the website to block page access when it detects an ad-blocker. It does so by changing the main content to a message requesting the user to remove their ad-blocker. As the network activity remains unchanged, there are no differences on features reflecting network requestsfor the broken sub-tree. Thus, a model trained only on network-based features would fails to detect this \textit{content} breakage. \name finds that the main content sub-tree has been shifted from its original position, has a large overall size, originally had one visible node, one \textit{Layout} nodes edited, it covers 2 salient elements (originally a video and some text), and the text content was changed. Using this, \name predicts that the subtree's edits cause breakage.

\subsection{Domain hopping and filter-list deprecation} 
\label{domain-hopping-example}
We say a webpage does \textit{domain hopping} if it redirects the user to different root domains across time (days or months). Websites employ domain hopping to evade any domain-based blocking or banning. For example, Adguard's issue \texttt{141090} presents a broken webpage formerly under the domain \texttt{gogohd.net}. During our investigation, the page automatically redirects to a new domain periodically (\eg \texttt{anihdplay.com}) and renders any domain-specific filter rule useless. Reproducing this breakage would require constantly updating filter rules, which is unmaintainable.

\subsection{Overlay breakage and \textit{cookie-banner} evasion limitations} 
\label{subsubsec:overlay-breakage-example}
Overlays are the background elements for full-page forms (\eg cookie consent, privacy policy, pay-wall \etc). They prevent the user from accessing the page before accepting the conditions in the form. 
Figure \ref{fig:example-overlay-breakage} (left) shows the grey overlay for the cookie form. 
Overlay breakage happens when the form is hidden and the user cannot accept to access the page. 
In some web-pages, the overlay can be transparent; the site appears normal but prevents interactions. Overlay breakage does not necessarily have to be static breakage. 

During a crawl, before we take a site snapshot, we search for any cookie banner and try to accept the cookies. 
%
%
Since the \textit{cookie-banner} evasion module implements a keyword-based approach to find the cookie banner and accept it, it fails to account for some languages and phrasings of the cookie banner  (Figure \ref{fig:overlay-breakage-failed} (left)). 
When this happens, the overlay is present in the visit with no filter lists $C_N$ and the visit with breaking rules $C_B$. 
According to the labeling rule in Figure \ref{fig:subtree-labeling}, it is labeled as neutral not broken.

\section{Implementation Details}

\subsection{Extracting URLs from forum posts.}
EasyList forum issue titles often contain the broken webpage's domain. We extract the post body text using BeautifulSoup4 4.11.1 and extract all the URLs within using standard URL regex. Then, we keep the URLs that have the same domain as the title. When we obtain have many URLs or none at all, we flag the issue for manual investigation and extract the test URL manually. 
For uBlock's forum, the breakage URL is often located after the text ``URL address of the web page'' or ``\#\#\# URL(s) where the issue occurs'', or before ``\#\#\# Category''. We compare the extracted URLs with the domain usually found in the issue's title.
For Adguard's forum, the breakage URL is well-structured -- it is present after the statements ``\#\#\# Issue URL'', ``** Issue URL'', or ``Where is the problem encountered?''. We drop all issues that do not follow this format. 

\subsection{VIPS Python Implementation} 
\label{subsubsec:vips-implementation}

\begin{figure}[!t]
	\centering
 \label{fig:vips-seg}
	\includegraphics[width=0.5\linewidth]{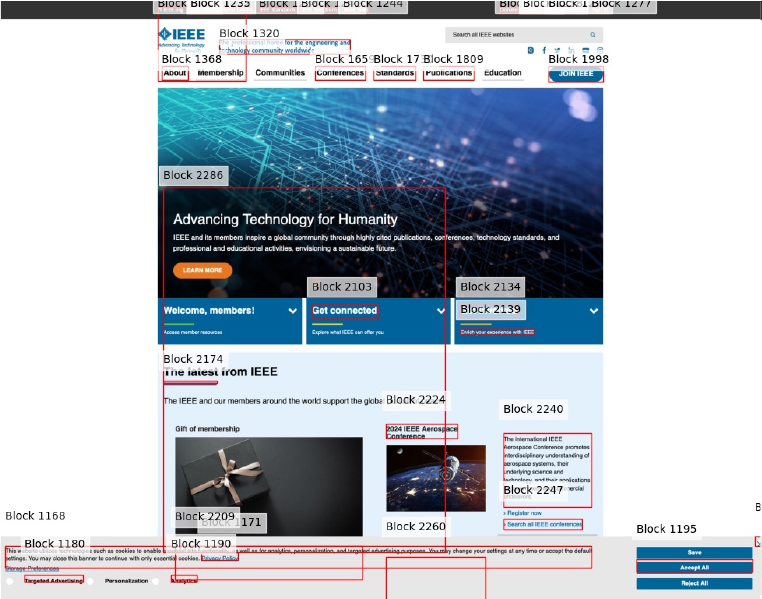}\hfill
 \includegraphics[width=0.5\linewidth]{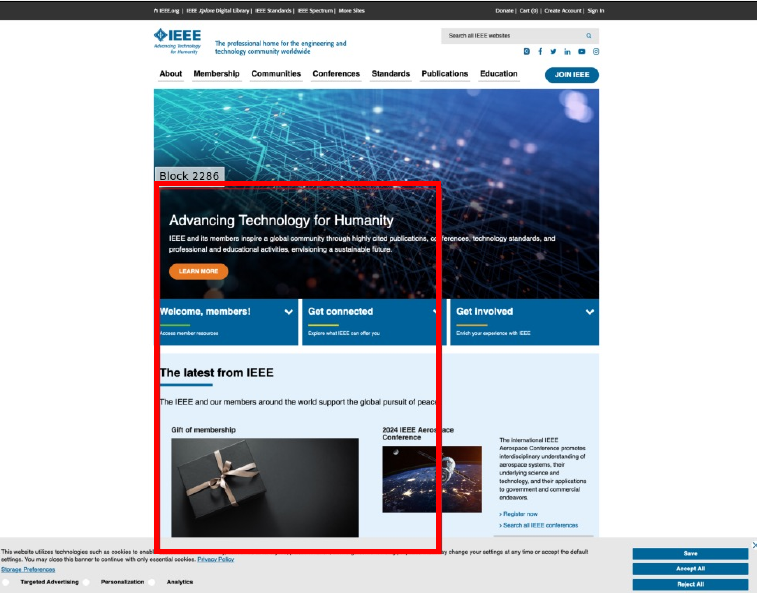}
	\caption[]{(top) Example of the output of VIPS Segmentation on a webpage. Each red rectangle represents one semantic group. (bottom) Example of the saliency prediction for the best model on the same website. The red rectangle represents the only salient group.}
\end{figure}

VIPS segments a page from a top-down iterative approach. 
In each iteration, it tries to find the optimal place to add a horizontal or vertical separator if any are possible. This divides larger segments into smaller ones. 
In this case, the round number is a hyperparameter that controls the granularity of this segmentation, since further rounds mean finer subdivisions of larger segments.  
VIPS relies on DOM and CSS heuristics to estimate the best location of a separator. For example, the boundaries of text with a similar font, the borders of tables, or a section with a different background color \etc. 
As Akpinar \etal~\cite{akpinar2013VisionBasedPage} points out, the original VIPS limitations arise due to changes by moving from HTML3 to HTML5. 
This change introduces tags and web-design practices previously unaccounted for by VIPS. For instance, webpages used to rely on \texttt{<br/>} to divide content but today it is rarely the case. Rather, we use \texttt{<div>...</div>} to structure the website. 
So, we updated a VIPS Python implementation ~\cite{wushuartgaro2023VipsPython} to account for these tags as candidates for separators. We also fine-tune internal hyperparameters by manually inspecting the results. 
First, we replace elements that represented dividers like \texttt{<hr/>} with \texttt{<div>} elements. 
Second, we add rules for many new elements that often represent media content, \eg \texttt{<object>}, \texttt{<embed>}, and \texttt{<iframe>}. 
Finally, VIPS focused on text-centric rules that are less relevant in current media-rich (image and graphics) websites, we change this focus by giving more weight to images and videos.
%

\subsection{Tree Comparison Algorithm}
\label{subsubsec:similarity-heuristic}
To compare two DOM trees $T_A$ and $T_B$, we start by comparing the children of the two roots, identifying children that are common and the same, common children that have attributes changed from $A$ to $B$, children that are unique to $T_A$, and those unique to $T_B$. If a node $x$ is unique to either tree, the sub-tree rooted at $x$ becomes a differential sub-tree (\textbf{added} or \textbf{removed}). If it is a common node with attributes changed we mark it as a root for an \textbf{edited} differential sub-tree. Finally, if the nodes are the same, they are part of the common tree $T_{A,B}$. We then repeat the algorithm above for the trees rooted at edited or similar nodes in both trees. 

\parait{Node similarity heuristic.} To compare children of $T_A$'s root and $T_B$'s root, we cannot simply use the DOM id, because it is generated at runtime and the browser might load elements in a different order in $A$ or $B$. We need to rely on \textit{attributes} and \textit{visual cues} for the nodes to determine whether they are the same. So, we implement a heuristic that computes a similarity score between nodes $a\in T_A$ and $b\in T_B$: $\text{sim}(a,b)\in [0,1]$. We then find the closest match to $a$ in $T_B$. 
\[
    \text{match}_{T_B}(a) := \operatornamewithlimits{arg\,max}_{b \in T_B} \text{sim}(a, b)
\]

Then, if the maximum score is between 0 and 0.75, we consider that $a$ was \textbf{removed}. If the maximum score is between 0.75 and 1, we consider that $a$ was \textbf{edited} to $\text{match}_{T_B}(a)$. Finally, if the maximum score is exactly 1, we add $a=\text{match}_{T_B}(a)$ as part of the common tree. These thresholds are manually fine-tuned and we don't claim they are ideal. 
The heuristic itself is composed of two parts, disqualification conditions that return a zero score immediately, and a numerical average score over \textit{attributes} and \textit{visual cues} of the pair of nodes. First, we list disqualification conditions: the nodes have different HTML \texttt{id}, \texttt{src}, or \texttt{name} values.
For the numerical score, we check the similarity of text content, classes in common for the \texttt{class} attribute, and the distance on the screen between the two nodes. 

This heuristic has limitations that we found empirically. Mainly, a large number of \texttt{<svg/>} pairs that should be considered \textbf{edited} have a max score less than 0.75, i.e. one is considered \textbf{added} while the other \textbf{removed}. The main reason behind this is that if an \texttt{<svg/>} does not load, the position is (0,0). So, the distance between the two nodes is large enough to reduce the score below 0.75.

\begin{table}[h]
\centering
\caption{Feature importance for saliency classifier according to AUC Loss predictive power.}
\label{tab:saliency_feature_importance}

\resizebox{\linewidth}{!}{
\begin{tabular}{rllr}
\toprule
    Rank & Category & Feature  &   AUC Loss  \\
\midrule
1  & Content & \% of layout nodes in this group
& 0.025 \\
2  & Content & Total \# of HTML \texttt{class} attributes & 0.025 \\
3  & Content & \% of layout nodes from global layout nodes&  0.022 \\
4  &  Positional & Mean X coordinate across all groups &0.022 \\
5  & Visual & Width of this group & 0.022 \\
6  & Content & Whether this group has the \texttt{id} attribute & 0.017 \\
7  & Content & \% of text nodes in this group & 0.014 \\
8  & Content & Total \# of functional nodes globally  & 0.014 \\
9  & Content & \% of HTML \texttt{class} attributes & 0.013 \\
10  & Content & Total \# of layout nodes in this group & 0.013 \\
11 & Positional & X coordinate of the center of this group & 0.013 \\
12 & Positional & Y coordinate of the center of this group & 0.012 \\
13 & Content & Total \# of layout layout nodes in all groups & 0.011 \\
14 & Content & Text content total entropy in this group & 0.010 \\
15 & Content & Classes values total entropy in this group & 0.010 \\
16 & Content & Total \# of text nodes in all group & 0.007 \\
17 & Structural & Total \# of nodes in all groups & 0.006 \\
18 & Content & Total text length in this group & 0.005 \\
19 & Visual & Mean color vibrancy of elements in the group & 0.005 \\
20 & Structural & Total \# of nodes in this group & 0.004 \\
21 & Content & \% of functional nodes from global layout nodes & 0.003 \\
22 & Content & Total \# of layout nodes in this group & 0.003 \\
23 & Content & Total \# of text nodes in this group & 0.002 \\
24 & Visual & Height of this group & 0.002 \\
25 & Content & \% of functional nodes in this group & 0.001 \\
26 & Visual & Size of the group (height $\times$ width) & 0.000 \\
27 & Visual & Centrality of the group = $e^{-10 ((\bar x - 0.5)^2 + (\bar y - 0.5)^2)}$ & 0.000 \\
28 & Visual & Mean font size in the group & 0.000 \\
29 & Positional & Mean Y coordinate across all groups & 0.000 \\
30 & Content & \% of text nodes from global layout nodes & 0.000 \\
31 & Visual & Mean font-weight in the group & 0.000 \\
\bottomrule
\end{tabular}
}
\end{table}

\begin{table}[h]
\centering
\caption{Feature ablation results: Top features according to AUC Loss predictive power.\\Scope: S=Subtree, G=Global. Category: V=Visual, S=Structural, F=Functional, C=Content. Crawl: $F^-$=Crawl before filter list change, $F^{+}$=Crawl after filter-list change. }
\label{table:sinbad-feature-importance}
\begin{tabular}{cllp{2in}}
\toprule
AUC Loss &    Scope &    Cat. &                                                                                      Description \\
\midrule
 0.0110 &  S &      V &   Size of the subtree on screen at $F^-$ \\
0.0093 &  S &      V &    Number of salient elements in subtree at $F^-$ \\
0.0082 &  S &  S &  Average degree of subtree nodes \\
 0.0082 &   G &  F & Total number of requests added from $F^-$ to $F^+$ \\
 0.0061 &  S &  F &  $\Delta$ in the total number of elements queried by a script related to the sub-tree. \\
0.0060 &  S &  F & $\Delta$ in the total number of elements queried after interactions with the sub-tree. \\
0.0056 &   G &  F &  Total  of requests removed from $F^-$ to $F^+$ \\
 0.0049 &  S &     C & Number of tags considered \textit{Text} added in the subtree from $F^-$ to $F^+$ \\

 0.0047 &  S &  F & Total number of interactions with the sub-tree \\

  0.0043 &  S &     C & Number of tags considered \textit{Text} removed in the subtree after the filter rule edit \\

 0.0040 &   G &  F &  Total number of errors thrown by scripts removed from $F^-$ to $F^+$ \\

 0.0037 &  S &     C &          Number of tags considered \textit{Other} edited in the subtree from $F^-$ to $F^+$ \\

 0.0035 &  S &     C &          Number of tags considered \textit{Layout} added in the subtree from $F^-$ to $F^+$ \\

 0.0033 &  S &  F &                                Total number of requests called by scripts related to the sub-tree. \\

 0.0030 &  S &     C &                   Number of \texttt{<iframe>} tags added in the subtree from $F^-$ to $F^+$ \\

 0.0028 &  S &         - &                                                                Whether the subtree was removed \\

 0.0022 &  S &         - &                                                                 Whether the subtree was edited \\

 0.0021 &  S &     C &  Number of tags considered \textit{Input/Output} removed in the subtree from $F^-$ to $F^+$ \\

 0.0020 &  S &     C &           Number of tags considered \textit{Text} edited in the subtree from $F^-$ to $F^+$\\
 0.0019 &  S &         - &                                                                  Whether the subtree was added \\

 0.0018 &  S &      V & Number of salient elements edited in the subtree after the filter rule edit \\
 0.0016 &   G &  F &                                                               Total number of requests in the page \\
 0.0013 &   G &  F &                                                                Total number of scripts in the page \\
 0.0012 &  S &  F &  Total number of requests called by elements in the subtree \\
 0.0010 &  S &     C &         Number of tags considered \textit{Layout} edited in the subtree from $F^-$ to $F^+$ \\

 0.0009 &  S &     C &         Number of tags considered \textit{Other} removed in the subtree from $F^-$ to $F^+$ \\

 0.0007 &  S &     C &   Number of tags considered \textit{Input/Output} edited in the subtree from $F^-$ to $F^+$ \\

 0.0006 &  S &     C &                 Number of \texttt{<iframe>} tags removed in the subtree from $F^-$ to $F^+$ \\
 0.0006 &  S &     C &                  Total number of \texttt{<iframe>} tags in the subtree from $F^-$ to $F^+$ \\

 0.0005 &  S &  F &                       $\Delta$ in the total number of errors after interactions with the sub-tree. \\

 0.0005 &  S &      V &                       Number of salient elements removed in the subtree from $F^-$ to $F^+$ \\
\bottomrule
\end{tabular}
\end{table}

\newpage
\hfill

\newpage

\section{Meta-Review}

The following meta-review was prepared by the program committee for the 2024
IEEE Symposium on Security and Privacy (S\&P) as part of the review process as
detailed in the call for papers.

\subsection{Summary}
Web privacy tools and ad blockers can cause websites to "break," which is when non-advertising, user-desired functions of a website no longer work due to those tools. This paper presents and evaluates SINBAD, a method for automatically detecting broken webpages via a classifier that analyzes subtrees on a webpage, finding roughly a 20\% improvement in accuracy over a partial re-implementation of the previous state of the art. This classifier was trained on differential web crawls in which the researchers visited pages with and without privacy tools and ad blockers enabled, successfully capturing aspects of the web's nondeterminism as well as modeling the most salient parts of pages to prioritize automated tests that approximate a user's interactions.

\subsection{Scientific Contributions}
\begin{itemize}
\item Creates a New Tool to Enable Future Science
\item Addresses a Long-Known Issue
\item Provides a Valuable Step Forward in an Established Field
\item Establishes a New Research Direction
\end{itemize}

\subsection{Reasons for Acceptance}
\begin{enumerate}
\item The training of the SINBAD tool and evaluation thereof follow a clear and systematic methodology.
\item The saliency-focused crawl appears to be a novel idea and crucially is not computationally heavy.
\item The differential approach in the crawl helps distinguish broken pages from the nondeterminism of the web.
\item The researchers created an interesting evaluation dataset based on user-reported issues on forums.
\item The paper reports a low false positive rate for SINBAD.
\item The ability to partially mitigate broken webpages is important for the adoption of privacy-enhancing tools for the web.
\end{enumerate}

\subsection{Noteworthy Concerns}
\begin{enumerate} 
\item The published code for the most related prior approach, by Smith et al., has missing and deprecated dependencies. Thus, this paper could not compare the SINBAD approach directly. Thus, they compare both against a partial re-implementation and against previously reported results from the Smith et al. paper from an earlier time, making it hard to separate artifacts of the different datasets and different implementations.

\item While the paper reports SINBAD's feature importance, without an ablation study and more precise comparison to prior work it is not possible to fully attribute the accuracy improvements that were shown to specific design decisions.

\item The dataset is small (i.e., not comprehensive) and may not be representative of all websites.

\item The evaluation set is unbalanced (overrepresenting broken pages), leading to questions about the false positive rate in an open world scenario. This concern has been partially addressed by a new experiment that measures the false positive rate on a small sample of websites, but the small size of that sample still leaves open questions about how these results will generalize.

\end{enumerate}

\section{Response to the Meta-Review} 

    \para{1)} We acknowledge the limitations of running a comparison between our re-implementation of Smith \etal's work \cite{smith2022blocked}. Yet, as \name's improvement is consistent across datasets (Table~\ref{tab:external-datasets}), we argue that potential artifacts of the datasets have no influence on our conclusions.

    \para{3 and 4)} We agree that given the size of the dataset, it may not be representative of all kinds of breakage on the Web. We would like to note that this limitation is inherent to having a quality dataset which requires manual validation. We made this choice, given that our checks on examples collected via heuristics~\cite{smith2022blocked} were inaccurate (see Section~\ref{subsec:compare-methods}). We would like to encourage the community to research better methods of collecting reliable data so that this issue can be addressed in a systematic manner.

\clearpage
\onecolumn

\end{document}